\documentclass[lettersize,journal]{IEEEtran} 
\IEEEoverridecommandlockouts

\usepackage[numbers]{natbib}
\usepackage[hyphens]{url}
\usepackage{hyperref}
\usepackage{amsmath,amssymb,amsfonts}
\usepackage{algorithmic}
\usepackage{textcomp}
\usepackage{xcolor}
\usepackage{amsmath}
\usepackage{algorithm}
\usepackage{subcaption}
\usepackage{wrapfig}
\usepackage{ragged2e} 

\usepackage[nolist,nohyperlinks]{acronym}  
 
\usepackage{tabularx}

\newcolumntype{R}{>{\raggedleft\arraybackslash}X}
\newcolumntype{L}{>{\raggedright\arraybackslash}X}
\newcolumntype{T}{>{\centering\arraybackslash}X}  
\newcolumntype{C}{>{\centering\arraybackslash}X} 

\usepackage{pdflscape,array}
\usepackage{graphicx,url}
\usepackage{makecell}
\usepackage{multirow}
\usepackage{pgf} 
\usepackage[super]{nth}
\usepackage{booktabs}

\usepackage{colortbl}
\definecolor{grayrow}{gray}{0.9} 
 
\begin{document} 
 
\title{TII-SSRC-23 Dataset: Typological Exploration of Diverse Traffic Patterns for Intrusion Detection}

\author{Dania\,Herzalla,~Willian\,T.\,Lunardi,~and~Martin\,Andreoni\,Lopez%

\IEEEcompsocitemizethanks{\IEEEcompsocthanksitem Dania Herzalla, Willian T. Lunardi, and Martin Andreoni Lopez are with the Technology Innovation Institute, 9639 Masdar City, Abu Dhabi, UAE -- \texttt{\{dania.herzalla, willian.lunardi, martin.andreoni\}@tii.ae}
}}

\markboth{Preprint submitted to IEEE, August 15, 2023}{} 

\maketitle

\begin{abstract}
The effectiveness of network intrusion detection systems, predominantly based on machine learning, are highly influenced by the dataset they are trained on. 
Ensuring an accurate reflection of the multifaceted nature of benign and malicious traffic in these datasets is paramount for creating IDS models capable of recognizing and responding to a wide array of intrusion patterns. However, existing datasets often fall short, lacking the necessary diversity and alignment with the contemporary network environment, thereby limiting the effectiveness of intrusion detection. This paper introduces TII-SSRC-23, a novel and comprehensive dataset designed to overcome these challenges. Comprising a diverse range of traffic types and subtypes, our dataset is a robust and versatile tool for the research community. Additionally, we conduct a feature importance analysis, providing vital insights into critical features for intrusion detection tasks. Through extensive experimentation, we also establish firm baselines for supervised and unsupervised intrusion detection methodologies using our dataset, further contributing to the advancement and adaptability of IDS models in the rapidly changing landscape of network security. Our dataset is available at \url{https://kaggle.com/datasets/daniaherzalla/tii-ssrc-23}.
\end{abstract}


\begin{acronym}
    \acro{PCAP}{Packet Capture}
    \acro{CSV}{Comma-Separated Values} 
    \acro{IDS}{Intrusion Detection System}
    \acro{RF}{Random Forest} 
    \acro{t-SNE}{t-distributed Stochastic Neighbor Embedding} 
    \acro{PFI}{Permutation Feature Importance} 
    \acro{XGBoost}{eXtreme Gradient Boosting} 
    \acro{ET}{Extra Trees}
    \acro{OOD}{Out-of-Distribution}
    \acro{SMOTE}{Synthetic Minority Over-sampling Technique}
    \acro{AUROC}{Area Under the Receiver Operating Characteristic Curve}
    \acro{AUC-PR}{Area Under the Precision-Recall Curve}
    \acro{MLP}{Multilayer Perceptron}
    \acro{SVM}{Support Vector Machine}
    \acro{DT}{Decision Tree}
    \acro{OC-SVM}{Support Vector Machine}
    \acro{IF}{Isolation Forest}
    \acro{KDE}{Kernel Density Estimator}
    \acro{LOF}{Local Outlier Factor}
    \acro{Deep SVDD}{Deep Support Vector Data Description}
    \acro{DoS}{Denial of Service}
    \acro{DDoS}{Distributed Denial of Service}
    \acro{MTU}{Maximum Transmission Unit}
    \acro{ML}{Machine Learning}
    \acro{DL}{Deep Learning}
    \acro{BATMAN}{Better Approach to Mobile Ad-hoc Networking}
    \acro{VLC}{Video LAN Client}
    \acro{IP}{Internet Protocol}
    \acro{UDP}{User Datagram Protocol}
    \acro{RTP}{Real-Time Transport Protocol}
    \acro{TS}{Transport Stream}
    \acro{HTTP}{Hypertext Transfer Protocol}
    \acro{PCM}{Pulse-Code Modulation} 
    \acro{TCP}{Transmission Control Protocol}
    \acro{FTP}{File Transfer Protocol}
    \acro{SSH}{Secure Shell Protocol}
    \acro{DNS}{Domain Name System}
    \acro{ICMP}{Internet Control Message Protocol}
    \acro{IoT}{Internet of Things}
    \acro{CnC}{Command-and-Control}
    \acro{SMTP}{Simple Mail Transfer Protocol}
    \acro{IMAP}{Internet Message Access Protocol}
    \acro{POP3}{Post Office Protocol}
    \acro{HTTPS}{Hypertext Transfer Protocol Secure}
    \acro{SIDDoS}{SQL Injection Dos}
    \acro{OS}{Operating System}
    \acro{IGMP}{Internet Group Management Protocol}
    \acro{ARP}{Address Resolution Protocol}
    \acro{RARP}{Reverse Address Resolution Protocol}
    \acro{IIoT}{Industrial Internet of Things}
    \acro{RTSP}{Real Time Streaming Protocol}
    \acro{Mbps}{Mega bits per second}
    \acro{CM4}{Compute Module 4}
    \acro{SCTP}{Stream Control Transmission Protocol}
\end{acronym}

\section{Introduction} \label{sec:introduction}
 
As the digital world becomes increasingly interconnected, the need for robust network security has become paramount. This increasing interconnectedness, driven by technologies ranging from mobile computing to the \ac{IoT}, brings with it an exponentially growing attack surface, making network security not merely an optional layer but a critical necessity. At the heart of this defense strategy lie \ac{IDS}. These systems employ many techniques, from statistical anomaly detection to signature-based methods and, increasingly, \ac{ML} approaches, to identify and mitigate anomalous or malicious activity within a network. When discussing the role of \ac{ML} in IDS, it's crucial to highlight the concept of data diversity, illustrated by practices like data augmentation. Data augmentation is a common technique to introduce variability into the training data in training \ac{ML} models, particularly \ac{DL} methods. This technique can prevent models from overfitting specific patterns and instead promote the ability to generalize to unseen instances. Similarly, the value of data diversity extends to network traffic datasets used for training IDS models, as it can enrich the models' ability to identify a broader range of intrusion scenarios.

Despite the critical importance of data diversity, traditional network traffic datasets, which are frequently employed in shaping network security approaches, exhibit significant limitations, most notably a lack of variation within the category of malicious samples. Although these datasets were pioneering at the time of their creation, they are now constrained by outdated patterns, inherent biases, and obsolete features that do not accurately reflect the ever-evolving landscape of modern network traffic. The lack of diversity, particularly within the malicious class, limits the ability of IDS models trained on these datasets to generalize effectively to new, unseen intrusions commonplace in today's complex networks. The \ac{IoT} has added another layer of complexity to network traffic, with its unique data patterns and its inherent security challenges. Despite efforts to create IoT-specific datasets, many of these initiatives fail to capture the full spectrum of device interactions and the diverse range of potential intrusions that can occur in these settings. The heterogeneity of IoT networks, characterized by a vast array of interconnected devices with varying capabilities and vulnerabilities, amplifies the challenge of curating a representative dataset. Consequently, this presents an urgent call for creating more comprehensive and diverse datasets that better encapsulate the contemporary threats networked systems face. 

In this paper, we propose TII-SSRC-23, a new dataset designed to address the challenges outlined earlier. The dataset totals 27.5~GB and is bifurcated into two main categories: benign and malicious, encompassing eight distinct traffic types. These types are divided into 32 traffic subtypes: six benign and 26 malicious. Both the raw network traffic data, stored as \ac{PCAP} files, and the extracted features, presented in the form of \ac{CSV} files, are included in our dataset. Our methodology for dataset generation begins with defining the network topology, serving as the foundation for all subsequent interactions. This includes generating benign traffic that mimics typical network interactions across unique data types such as video, audio, text, and background traffic. Following this, we outline the generation of malicious traffic, replicating four types of network threats: \ac{DoS} attacks, brute-force attacks, information gathering tactics, and botnet traffic, with a specific emphasis on the Mirai botnet. Feature extraction and importance are analyzed, followed by supervised and unsupervised experiments that establish firm baselines for future works. Our main contributions can be summarized as follows:
\begin{itemize}
\item We present the open-source TII-SSRC-23 dataset, a heterogeneous collection encompassing eight traffic types (audio, background, text, video, bruteforce, \ac{DoS}, information gathering, botnet) and 32 subtypes across both benign and malicious categories. 
\item We conduct an exhaustive survey on 18 existing network traffic datasets, providing key insights to aid researchers in dataset selection for IDS research.
\item We perform a comprehensive feature importance analysis within network traffic data, offering valuable insights on critical features for intrusion detection tasks, thereby facilitating IDS model optimization.
\item Through extensive experimental evaluation, we establish firm baselines for supervised and unsupervised intrusion detection methodologies using our dataset, fostering the development of robust IDS systems optimized for diverse network traffic situations.
\end{itemize} 

The remainder of this paper is structured as follows: Section~\ref{sec:related} provides a comprehensive review and analysis of preceding work that centers around creating and publicly releasing network traffic datasets, tackling the limitations and challenges inherent to existing data sources. Section~\ref{sec:dataset} provides an exhaustive description of our proposed network IDS dataset generation process, encompassing the testbed, the types, and the characteristics of both benign and malicious traffic. In Section~\ref{sec:feat_engineering}, we examine statistical patterns and characteristics of the produced network traffic through the lens of feature importance analysis. This includes data preprocessing stages, feature extraction via CICFlowMeter~\cite{draper2016CICFlow}, and feature importance computations to discern the most informative features. Section~\ref{sec:experiments} is dedicated to evaluating both supervised and unsupervised methodologies to set solid baseline performances for intrusion detection using our dataset. Conclusively, Section~\ref{sec:conclusion} wraps up the paper.

\begin{table*}[t]
    \centering
    \caption{IDS Datasets Characteristics}
    \label{tab:ids_datasets_survey}
    \begin{tabularx}{\textwidth}{l>{\hsize=1.23cm}CCC>{\hsize=1.25cm}C>{\hsize=1.15cm}C>{\hsize=2.3cm}C>{\hsize=2.6cm}C} 
        \toprule
        \textbf{Dataset} & \textbf{Year} & \textbf{\# Traffic Objects} & \textbf{Published Format} & \textbf{Size (GB)} & \textbf{Features} & \textbf{Traffic Source} & \textbf{Testbed} \\
        \midrule
        DARPA98 & 1998 & -- & Raw & 4 & -- & Emulated & Small (military) \\ 
        \rowcolor{gray!10}
        KDD99 & 1998 & 4.9M bi. flows & Statistics & -- & 41 & Emulated & Small (military) \\
        NSL-KDD & 1998 & 1M bi. flows & Statistics & -- & 41 & Emulated & Small (military) \\
        \rowcolor{gray!10}
        Kyoto 2006+ & 2006-09 & 93M bi. flows & Statistics & -- & 24 & Real & Large (honeypots) \\ 
        UNIBS & 2009 & 79k bi. flows & Raw, statistics & 2.7 & 8 & Real & Medium (university) \\ 
        \rowcolor{gray!10} 
        CTU-13 & 2011 & 81M bi. flows & Raw, statistics & 77 & 14 & Real & Large (university) \\ 
        TUIDS & 2011-12 & 250k bi. flows & Raw, statistics & -- & 50, 24 & Real & Large (university) \\ 
        \rowcolor{gray!10} 
        ISCX 2012 & 2012 & 2M bi. flows & Raw, statistics & 84.1 & 14 & Emulated & Small \\ 
        UNSW-NB15 & 2015 & 2.5M bi. flows & Raw, statistics & 99.1 & 49 & Synthetic & Small \\ 
        \rowcolor{gray!10}
        DDoS 2016 & 2016 & 2.1M bi. flows & Statistics & -- & 27 & Synthetic & -- \\ 
        CICIDS 2017 & 2017 & 3.1M bi. flows & Raw, statistics & 47.9 & 80 & Emulated & Medium \\
        \rowcolor{gray!10}
        CIC DoS & 2017 & -- & Raw & 4.6 & -- & Emulated & Small \\
        N-Baiot & 2018 & 7M  & Statistics & -- & 115 & Emulated & Small (IoT) \\
        \rowcolor{gray!10} 
        BoT-IoT & 2019 & 73M bi. flows & Raw, statistics & 69.4 & 46 & Emulated, synthetic & Small (IoT) \\
        TON-IoT & 2019 & 22M bi. flows & Raw, statistics & 65.1 & 44 & Emulated, synthetic & Medium (IoT) \\
        \rowcolor{gray!10} 
        CIC IoT & 2022 & 30k bi. flows & Raw, statistics & 60.3  & 48 & Emulated & Medium (IoT) \\
        LATAM-DDoS-IoT & 2022 & 49M bi. flows & Raw, statistics & 279.8 & 20 & Real, emulated & Large (IoT) \\
        \rowcolor{gray!10}
        Edge-IIoTset & 2022 & 20M bi. flows & Raw, statistics & 69.3 & 61 & Emulated & Medium (IIoT) \\ 
        \midrule
        TII-SSRC-23 (ours) & 2022-23 & 8.6M bi. flows & Raw, statistics & 27.5 & 75 & Emulated & Small \\
        \bottomrule
    \end{tabularx}
\end{table*}

\section{Related Works} \label{sec:related}

In this section, we delve into a comprehensive timeline of IDS datasets spanning the last quarter-century, from earlier published datasets in 1998 to more recent ones released in 2023. We review a range of datasets including some of the more traditional testbed datasets featuring network-layer attacks, real-world network deployments, and IoT datasets. Table~\ref{tab:ids_datasets_survey} presents a survey of the datasets, considering characteristics such as the year of the dataset's creation, number of traffic objects, dataset's published format, size of the raw traffic, number of features extracted from the dataset, traffic source, and deployed network topology. The number of traffic objects is either represented as a value with the bidirectional flows~\footnote{Formal definitions of unidirectional and bidirectional network flows can be found in Appendix~\ref{sec:prob_notation}.} label (bi. flows) or just as a value. The latter implies that no information was found regarding the type of traffic object of the dataset. The published format, which represents the form in which the data was published, is described either as raw, denoting that the network traffic provides packet-level information or as statistics, providing information about the traffic objects. The traffic source falls into three categories: real, emulated, or synthetic. Real denotes that the data was captured in a real-world network deployment, emulated refers to the data being captured in a controlled network environment with traffic generated manually, and synthetic means a network traffic simulation tool was used to generate data. Finally, for the testbed, small indicates that the testbed contained few than 20 nodes, medium indicates that the testbed contained 20 to 50 nodes, and large implies that a real-world network deployment or a testbed consisting of more than 50 nodes was used. In the case that we could not find specific information for a dataset or is irrelevant considering the data available, it is indicated by a dashed mark. 

The DARPA98 dataset~\cite{mit1998DARPA} established a performance benchmark for intrusion detection systems with a military network testbed showcasing diverse traffic types like \ac{DoS}, probing, and privilege escalation attacks. This dataset inspired the development of the KDD99 dataset~\cite{stolfoKDD}, which processed the raw traffic portion of the DARPA98 dataset comprising of benign and malicious traffic. Despite its merits, KDD99 had a significant problem of redundant records~\cite{tavallaee2009NSLKDD}, leading to the inception of the NSL-KDD dataset~\cite{tavallaee2009NSLKDD}. NSL-KDD, a polished version of KDD99, underwent preprocessing to eliminate redundancy, offering a more realistic evaluation context for intrusion detection systems and anomaly detection algorithms. However, these datasets share a key limitation -- their outdatedness hinders their utility for modern network traffic analysis. The Kyoto 2006+ dataset~\cite{kyoto}, which encapsulates real-world network traffic data harvested from Kyoto University between 2006 and 2009 using honeypots, has its limitations. It lacks manual labeling and introduces anonymization, and its network traffic perspective is constrained to honeypot-targeted attacks.  While the dataset incorporates ten additional attributes compared to the aforementioned datasets that are useful for IDS investigation, the benign traffic simulation is limited to \ac{DNS} and mail traffic data, excluding a more extensive range of real-world benign traffic.

The ISCX 2012 dataset~\cite{shiravi12iscx} used an innovative approach involving $\alpha$ and $\beta$ profiles to mimic benign user activities and malicious scenarios. The benign user behavior included traffic from the protocols: \ac{HTTP}, \ac{SMTP}, \ac{SSH}, \ac{IMAP}, \ac{POP3}, and \ac{FTP}. This dataset includes raw packet-level data in \ac{PCAP} files, featuring approximately 2.4 million bidirectional flows. Echoing this methodology, the CICIDS2017 dataset~\cite{sharafaldin2018CICIDS} generated a realistic background traffic scenario using the B-Profile system. This system models the behavior of 25 users based on \ac{HTTP}, \ac{HTTPS}, \ac{FTP}, \ac{SSH}, and email protocols. It comprises six attack profiles, specifically bruteforce, heartbleed botnet, \ac{DoS}, \ac{DDoS}, web, and infiltration attacks. Developed in 2015, the UNSW-NB15 dataset~\cite{moustafa15UNSW} comprises benign and malicious network traffic data generated using a network traffic simulation tool over a week in a controlled setting. The dataset includes nine attack classes: backdoors, \ac{DoS}, exploits, fuzzers, and worms. Presented in packet-based format (\ac{PCAP}) and bidirectional flow-based format, it features 49 attributes and predefined train-test splits. The dataset contains around 2.5 million bidirectional flows with an estimated 2.8\% malicious traffic. The UNIBS dataset~\cite{gringoli2009UNIBS} consists of traffic collected on the edge router of a campus network using 20 workstations. The traffic collected provides valuable network traffic information related to the campus network's communication patterns and behavior. However, the dataset does not contain malicious traffic traces. The CTU-13 (Capture The Flag) dataset~\cite{garcia2014CTU} contains real botnet traffic mixed with benign traffic captured in a university network. The malicious traffic includes 13 scenarios of botnet samples in which each scenario included botnet, benign, C\&C, and background flows. The dataset is labeled to indicate the type of malware attack. It is available in \ac{PCAP} and bidirectional flow-based format. The TUIDS dataset~\cite{bhuyan2015TUIDS} encompasses benign user behavior and various malicious traffic types including botnet, \ac{DoS}/\ac{DDoS}, probing, coordinated port scan, and privilege escalation. The data was generated using approximately 250 clients. The dataset, captured in raw packet-level and bidirectional flow formats, is labeled and contains around 250k flows. As the dataset is not publically available, we could not determine the size of the raw traffic. Shifting the focus to \ac{DoS}- and \ac{DDoS}-based datasets, the \ac{DDoS} 2016 dataset~\cite{alkasassbeh2016DDoS} contains benign traffic instances and focuses on \ac{DDoS} attacks such as \ac{UDP} flood, smurf, \ac{HTTP} flood, and \ac{SIDDoS}. However, the traffic was generated using a network traffic simulator. The CIC DoS dataset~\cite{jazi2017CICDoS} focuses on eight different application layer \ac{DoS} attacks, particularly HTTP \ac{DoS}. To create benign traffic that mimics normal user behavior, traffic from the ISCX 2012 dataset was used. The dataset is provided in raw capture format, making it useful for studying and evaluating intrusion detection methods in the context of application layer HTTP \ac{DoS} attacks.

As for \ac{IoT}-based datasets, the BoT-IoT dataset~\cite{moustafaBotIoT} offers a mix of benign and botnet traffic, simulating a realistic network environment. It comprises synthetically created benign traffic as well as diverse attack types such as \ac{DDoS}, \ac{DoS}, \ac{OS} and service scan, keylogging, and data exfiltration attacks, with \ac{DDoS} and \ac{DoS} attacks further classified by protocol. The dataset incorporates protocols like \ac{TCP}, \ac{UDP}, \ac{ARP}, \ac{ICMP}, \ac{IGMP}, and \ac{RARP}. The dataset features around 73 million bidirectional flows. The LATAM-DDoS-IoT dataset~\cite{almaraz2022LATAM} is designed with a primary focus on \ac{DoS} and \ac{DDoS} attacks, implemented in a testbed of physical and virtual \ac{IoT} components. Benign traffic from a production network was collected. The dataset includes two versions: LATAM-DoS-IoT and LATAM-DDoS-IoT, with 30 and 49 million bidirectional flows, respectively. The CIC IoT 2022 dataset~\cite{dadkhah2022CICIoT} was developed for the profiling, behavioral analysis, and vulnerability testing of IoT devices using various protocols. It collects data from experiments covering power-on, idle, interactions, scenarios, active network communications, and attack traffic: flood and \ac{RTSP} bruteforce. The collection process targeted IoT devices linked to an unmanaged switch, simulating a wireless IoT environment. The Edge-IIoTset dataset~\cite{ferrag2022EdgeIoT} caters to \ac{IoT} and \ac{IIoT} applications. The dataset is a multi-layered testbed, utilizing more than 10 different \ac{IoT} devices, and encompasses 14 attacks related to \ac{IoT} and \ac{IIoT} connectivity protocols. These attacks are categorized into five threats, including \ac{DoS} and \ac{DDoS}, information gathering, injection, man-in-the-middle, and malware attacks. The dataset contains around 20 million bidirectional flows, with about 11.2 million benign and 9.7 million malicious, with 61 extracted traffic features. The TON-IoT dataset~\cite{moustafa2021TONIoT} integrates \ac{IoT} and \ac{IIoT} systems and devices across edge, fog, and cloud layers within an orchestrated testbed architecture. The data encapsulates both synthetically created benign traffic and nine attack scenarios, shared as raw and processed traffic data in \ac{PCAP} and \ac{CSV} formats, along with operating system logs. The dataset comprises approximately 22.3 million bidirectional flows captured in 44 features. The benign traffic represents around 3.6\% of the flows in the dataset, leaving about 96.4\% as malicious flows. Lastly, the N-BaIoT dataset~\cite{meidan2018Nbaiot}, captured in an \ac{IoT} lab environment, records benign and botnet events. The dataset includes network traffic data from nine \ac{IoT} devices and encompasses 10 attack types originating from the BASHLITE and Mirai botnets. Featuring 23 distinct features, the dataset is shared in \ac{CSV} format. The dataset comprises over 7 million flows.

Table~\ref{tab:dataset_attacks} lists the attacks executed in all the aforementioned \ac{IDS} datasets. Although multiple datasets exist, such as UNSW-NB15 and CICIDS 2017, encompass many attack categories, our dataset concentrates on a wide breadth of each attack. Specifically, we investigate a variety of attacks within each of our four categories: \ac{DoS}, bruteforce, information gathering, and botnet. This investigation results in a total of 26 unique attacks launched.

\section{TII-SSRC-23: Dataset Generation Methodology}\label{sec:dataset}
In this section, we detail our methodology for creating the proposed 27.5 GB dataset in \ac{PCAP} format. The traffic is bifurcated into two primary categories (benign and malicious), spanning eight traffic types (audio, background, text, video, bruteforce, \ac{DoS}, information gathering, Mirai botnet), including 32 subtypes (six benign and 26 malicious). 
Table~\ref{tab:dataset_statistics} identifies the traffic types and subtypes, with each subtype quantified by the number of combinations\footnote{The number of combinations can exceed the number of bidirectional flows; this strictly depends on the protocol and how they are terminated.} and bidirectional flows. Moreover, the ``combinations" column denotes the traffic variations within a traffic subtype, approximated by the number of traffic permutations launched informed by the subtype's parameters, as listed in Appendix Section~\ref{sec:appendix_diversification}. 
The necessity for diversifying traffic patterns to enhance the resilience of \ac{IDS} is examined in Section~\ref{sec:diversification}. Our methodology begins with the specification of the network topology, outlined in Section~\ref{sec:testbed}, which forms the foundation for all subsequent interactions. The generation of benign traffic, emulating typical network interactions across the following unique data types: video, audio, text, and background traffic, is illustrated in Section~\ref{sec:ben_traffic}. Finally, Section~\ref{sec:mal_traffic} describes the generation of malicious traffic, replicating four types of network threats. 

\begin{table}
    \centering
    \caption{Distribution of bidirectional network traffic flows in the dataset, classified by type and subtype.}\label{tab:dataset_statistics}
    \setlength{\leftmargini}{0.3cm}    
    \begin{tabularx}{\columnwidth}{@{}>{\hsize=0.5cm}X>{\hsize=1.3cm}XC>{\hsize=0.3cm}RR@{}} 
        \toprule
        \textbf{Cat.} & \textbf{Type} & \textbf{Subtype} & \textbf{Combinations} & \textbf{Bi. Flows} \\
        \midrule
        \multirow{6}{*}{\rotatebox[origin=c]{90}{Benign}} & Audio & Audio & 1 & 190 \\
        \cmidrule{2-5}
         & Background & Background & 1 & 32 \\
        \cmidrule{2-5}
         & Text & Text & 1 & 209 \\
        \cmidrule{2-5}
         & \multirow{3}{*}{Video} & HTTP & 180 & 376 \\
        & & RTP & 180 & 349 \\
        & & UDP & 180 & 145 \\
        \midrule
        \multirow{26}{*}{\rotatebox[origin=c]{90}{Malicious}} & \multirow{5}{*}{Bruteforce} & \ac{DNS} & 2 & 22179 \\
        & & FTP & 1 & 3485 \\
        & & HTTP & 2 & 628 \\
        & & SSH & 1 & 3967 \\
        & & Telnet & 1 & 4913 \\
        \cmidrule{2-5}
        & \multirow{12}{*}{DoS} & ACK & 24 & 936307 \\
        & & CWR & 24  & 872523 \\
        & & ECN & 24 & 871150 \\
        & & FIN & 24 & 725600 \\
        & & HTTP & 27 & 82351 \\
        & & \ac{ICMP} & 16 & 9 \\
        & & MAC & 1 & 30 \\
        & & PSH & 24 & 909507 \\
        & & RST & 24 & 1072504 \\
        & & SYN & 24 & 856764 \\
        & & UDP & 24 & 257994 \\
        & & URG & 24 & 906190 \\
        \cmidrule{2-5}
        & Information Gathering & Information Gathering & 102 & 1038363 \\
        \cmidrule{2-5}
        & \multirow{8}{*}{Mirai} & DDoS ACK & 3 & 3779 \\
        & & DDoS \ac{DNS} & 1 & 55196 \\
        & & DDoS GREETH & 6 & 43 \\
        & & DDoS GREIP & 6 & 49 \\
        & & DDoS HTTP & 8 & 8923 \\
        & & DDoS SYN & 12 & 14210 \\
        & & DDoS UDP & 6 & 71 \\
        & & Scan and Bruteforce & 1 & 8731 \\
        \bottomrule
    \end{tabularx} 
\end{table}

\subsection{Traffic Diversification for Improved IDS Robustness}\label{sec:diversification}
Despite the impressive performance of various IDS datasets evaluated through \ac{ML}/\ac{DL} methodologies within their corresponding test environments, a significant performance decline is observed when these models are implemented in real-world contexts~\cite{sommerMLIDS}. This performance degradation often results in expensive misclassifications due to high false positive or false negative rates, thereby underlining a predominant challenge encountered by \ac{ML}-driven \acp{IDS}. An effective mitigation strategy involves utilizing network traffic datasets with diversified characteristics during training. This diversification allows the models to generalize better and accurately classify network traffic in real-world deployments. Although numerous existing datasets underscore the incorporation of an extensive variety of benign and malicious traffic, the emphasis on including diverse traffic patterns within each traffic category is noticeably lacking. In contrast, our proposed IDS dataset adopts a unique approach by stressing the generation of diversified traffic patterns within each traffic category. This is achieved through carefully manipulating data traffic parameters during the data generation stage, as described in the subsequent sections. By integrating this degree of diversity, our dataset is designed to enhance the robustness and effectiveness of \ac{ML}-based \acp{IDS}, particularly when facing an array of complex and evolving network traffic situations.

\subsection{Network Configuration Overview}\label{sec:testbed}
Our data recording setup captured benign and malicious traffic, deploying a testbed configuration composed of five nodes. These nodes encompass two laptop systems running Ubuntu 20.04 and three embedded devices, each offering processing capabilities equivalent to a Compute Module 4 device. Two of the embedded devices are interconnected to each laptop via Ethernet connections. At the same time, the third embedded device operated as a mobile unit, allowing placement in various locations, thus facilitating the simulation of diverse network interference scenarios. During traffic recording, the mobile embedded device is strategically relocated across three distinct locations to generate variations in network interference. The labels ``low," ``mid," and ``high" interference, which are relative terms, denote the distinct degrees of interference experienced at each respective location, as determined by the corresponding throughput values, i.e., approximately 38.4 \ac{Mbps} for ``low," 69.7 \ac{Mbps} for ``mid," and 154 \ac{Mbps} for ``high" interference scenarios. Specifically, at the first location the mobile device is placed half a meter away from the testbed, leading to the lowest level of interference. At the second location, the mobile device is stationed six meters horizontally away from the testbed, separated by two rooms, resulting in the highest level of interference. In contrast, the third location sees the mobile device placed six meters below the testbed, precisely one floor apart, creating mid-level interference. During all traffic capture scenarios, the tcpdump tool\footnote{Tcpdump: Unix-based network packet analyzer \url{http://www.tcpdump.org/}} was set to capture the traffic on the mobile embedded device. The embedded devices operate within a decentralized system where peer-to-peer communication occurs via a wireless medium. The traffic flow path is managed via the \ac{BATMAN} protocol chain, maintaining a static bi-directional path. This setup ensures that the communication passes through all nodes within the \ac{BATMAN} chain before reaching the destination node. 

In the Mirai malware attack scenario, the communication between the \ac{CnC} server and the bots does not follow an end-to-end path. Consequently, to comprehensively capture all \ac{CnC} and botnet traffic we recognized the need to construct a centralized testbed. This modified testbed included five nodes, with a Raspberry Pi 4 set as the Access Point, two Ubuntu 20.04 laptop systems as the victim and the botmaster hosting the \ac{CnC} server, and the ScanListen server, as well as two bots deployed on \ac{CM4} boards. All traffic was recorded on the Access Point using the tcpdump tool to capture bidirectional communication between the botmaster, the bots, and the victim.

\subsection{Benign Traffic} \label{sec:ben_traffic}

Our data collection, within the context of benign traffic, comprises four distinct types: audio, video, text, and background. Video traffic comprises the majority of benign flows, accounting for more than 65\% as deduced from Table~\ref{tab:dataset_statistics}. Audio and text traffic each comprises around 15\%, with background traffic making up around 3\%.

\subsubsection{Audio and Text Traffic}\label{sec:audio_text}
The Mumble\footnote{Mumble: open-source voice chat application. \url{https://www.mumble.info/}} voice-over \ac{IP} application was utilized to create audio and text traffic independently. The interaction between the client and the server was enabled using the Pymumble Python module, with a Python script devised to transmit audio and text messages using a script with over 100 varied-length strings. The network environment incorporated one server and three clients. The server was set on an embedded device, and the two laptop machines operated as clients, transmitting messages to the server. The clients dispatched audio/text messages with a 5\% probability of disconnection from the server. Upon disconnection, the client system was programmed to automatically re-establish the connection after a brief intermission. The audio and text traffic was each captured over a period of one hour. 


\subsubsection{Background Traffic}\label{sec:background}
The background traffic was recorded for a span of one hour. This strategy was twofold: not only did it contribute to the dataset by gathering background traffic, but it also provided a reference framework to aid in manually identifying specific background data types requiring filtration from the attack \ac{PCAP} files.

\subsubsection{Video Traffic} \label{sec:video}
The \ac{VLC} application was employed to generate video traffic, leveraging its accompanying Python module for automated video streaming. A custom Python script was created to introduce heterogeneity in the video traffic by modulating ten distinct video streaming parameters: pixel resolution, video codec, audio codec, video bitrate, audio bitrate, video scale, frames per second, multiplexer type, sample rate, and the underlying protocol. The \ac{VLC} streaming server was instantiated on the laptop. This server was responsible for streaming a playlist of seven unique videos. The streaming session was allotted one hour, where protocols such as \ac{UDP}, \ac{RTP}/\ac{TS}, and \ac{HTTP} were utilized for transmission. This procedure led to the creation of a \ac{PCAP} file for each of the utilized communication protocols. Comprehensive details regarding the modulated video traffic parameters are available in Appendix Table~\ref{tab:ben_params}.

\subsection{Malicious Traffic} \label{sec:mal_traffic}
In the context of malicious traffic, our compiled dataset embodies four different attack types. These comprise \ac{DoS}, Bruteforce, Information Gathering, and Botnet. The \ac{DoS} attacks represent the majority, accounting for approximately 86\% of the malicious traffic flows, followed by Information Gathering accounting for 12\%. Mirai Botnet and bruteforce each constitute 1\% of the malicious traffic.  After the data capture, a filtering process was applied to the attack \ac{PCAP} files during preprocessing, purging them of non-malicious data, as expanded upon in Section~\ref{sec:preprocessing}.  


\subsubsection{DoS}\label{sec:dos}
\ac{DoS} attacks, regarded as one of the most pervasive and frequently exploited types of network traffic intrusions, have witnessed a surge in both frequency and intensity in recent years. The year 2015 was a notable milestone in the history of \ac{DoS} attacks, setting unprecedented records for data flood transfer rates, a trend that intensified in the following year~\cite{macia2010defense}. These attacks, infamous for their disruptive effects, can rapidly deplete their targets' computational resources and bandwidth within minutes, effectively denying access to legitimate users. Reflecting the significant relevance of these attacks and in line with this trend, more than 85\% of our dataset constitutes \ac{DoS} attacks. Our investigation covers 12 unique flood attacks, each exploiting distinct vulnerabilities to inundate target devices. These attacks span HTTP, \ac{ICMP}, MAC, \ac{TCP} (ACK, CWR, ECN, FIN, PSH, RST, SYN, URG), and UDP. To incorporate variability and diversify the traffic, we meticulously modulated multiple parameters during the deployment of these attacks. Parameters such as speed of packet transmission and payload size can be key indicators of a \ac{DoS} attack.~\cite{alkasassbeh2016DDoS} Given their significance in the identification of \ac{DoS} activities, we dedicated special attention to manipulating them in order to capture the various ways these parameters are exploited by attackers. Within the \ac{ICMP}, \ac{TCP}, and \ac{UDP} floods we adjusted the speed of packet transmission to three distinct modes specified in Hping3: ``fast", ``faster", and ``flood", ranging from 10 packets per second (pps) to over 1000 pps, capturing a range of stealthy to aggressive flood attacks. Additionally, we varied the payload size to range from small inconspicuous payloads to larger payload flooding tactics to capture diverse DoS attack strategies.


The \ac{TCP} flood attack capitalizes on the intrinsic features and behavior of the \ac{TCP} protocol, exploiting the interactions using various flags present within the \ac{TCP} packets. As listed above, we launched eight distinct types of \ac{TCP} flood attacks. Within each, we varied six attack-related parameters: packet transmission speed, payload size, randomized source ports, \ac{TCP} checksum validity, \ac{TCP} window size, and \ac{TCP} data offset. This yielded 192 unique \ac{TCP} flood traffic combinations captured over 18.7 minutes. The UDP flood attack operates by transmitting a large volume of UDP packets. We modulated four parameters: packet transmission speed, payload size, randomized source ports, and UDP checksum validity, producing eight UDP traffic combinations captured over a period of three minutes. In the case of the \ac{ICMP} flood, we varied the payload size resulting in four unique combinations of traffic captured over a period of two minutes.  The \ac{HTTP} flood attack is a type of volumetric application layer attack that aims to inundate the target with \ac{HTTP} requests. We modulated three parameters for this attack: request method (GET, POST, Random), number of concurrent workers, and number of concurrent sockets. This configuration resulted in 27 unique traffic streams spanning a period of 11.3 minutes. The MAC flood was launched for 30 minutes, with no parameters adjusted, as the macof tool does not provide any traffic options to vary. In Appendix Table~\ref{tab:attack_params}, we provide further details of the modulated parameters for each flood attack, offering deeper insights into the experimental setup and configuration for our \ac{IDS} dataset.

\subsubsection{Bruteforce}\label{sec:bruteforce}
Despite their age and lack of sophistication, bruteforce attacks retain startling prevalence and efficacy in the contemporary digital landscape. This attack involves systematically attempting all possible combinations of credentials from a list of keys to discover a successful pair. The Patator tool\footnote{Patator: multi-purpose bruteforcer \url{https://www.kali.org/tools/patator/}} was used to execute bruteforce attacks on five services: \ac{DNS} (forward and reverse lookup), \ac{FTP}, \ac{HTTP}, \ac{SSH}, and Telnet. For launching the bruteforce attacks on the \ac{FTP}, \ac{HTTP}, \ac{SSH}, and Telnet services, we used a list of around 400k usernames and two million leaked passwords\footnote{The list of credentials used were obtained from a bruteforce database. \url{https://github.com/duyet/bruteforce-database/tree/master}}. The Filezilla Client application was set on the victim to perform the FTP bruteforce attack. The HTTP bruteforce attack was executed against a phpMyAdmin server hosted on the victim's machine, using GET and POST request methods. To carry out the forward \ac{DNS} lookup, we tested around 12k domain names against the server domain. The reverse \ac{DNS} lookup involved querying a range of \ac{IP} addresses to identify the victim's hostname.


\subsubsection{Information Gathering}\label{sec:info_gathering}
An information-gathering attack constitutes a critical initial step for attackers preparing for future exploits on their target system, proving particularly beneficial for malware attacks. Such an attack aims to acquire, among other things, information on a network's architecture, \ac{OS}, and active security defense mechanisms. Information-gathering attacks manifest in several forms, of which we implemented six types, specifically: port scan (\ac{TCP} and \ac{UDP}), \ac{OS} detection, version detection, script scan, and ping scan utilizing the Hping3 and Nmap\footnote{Nmap: open-source utility for network discovery and security auditing. \url{https://nmap.org/}} tools. We employed various \ac{IDS} evasion strategies to circumvent detection to render the scans more covert. 

A port scan involves scanning the ports of the victim to ascertain their status. The execution of a successful port scan provides the attacker with an entry point to penetrate the network and extract the targeted information. Hping3 was utilized to perform a scan on all ports using six \ac{TCP} flags. Additionally, Nmap was used to perform a \ac{UDP} scan and seven types of \ac{TCP} port scans with multiple parameters varied for each. The TCP port scans were of the following types: Connect, SYN ACK, FIN, Window, Maimon, XMAS, and NULL. As for a ping scan, it operates to discern the presence of hosts in a network by using their \ac{IP} addresses. Nmap was deployed to launch seven ping scans, namely: \ac{ICMP} echo, \ac{ICMP} timestamp request, \ac{ICMP} netmask request, \ac{TCP} SYN, \ac{TCP} ACK, \ac{UDP}, and \ac{SCTP} Initialization (INIT) scans. Finally, OS detection, version detection, script scanning, and traceroute techniques were performed. This was facilitated using the pre-configured ``Aggressive Scan" Nmap option, which activates multiple advanced scans to probe the target machine comprehensively. All of the information gathering tactics yielded 102 unique combinations of traffic, elaborated upon in Appendix table~\ref{tab:attack_params}.


\subsubsection{Botnet Malware}\label{sec:mirai}
In the field of cybersecurity, malware--a form of software-based attack--poses a significant threat by compromising system confidentiality. This breach can lead to sensitive data theft, disruption in system operations, or render the system entirely inoperative. Among various types of cyberattacks against embedded systems, botnet malware is one of the most prevalent~\cite{mirai}. The Mirai botnet is a notable example of this type of malware~\cite{mirai}. Designed specifically to infiltrate devices running a Linux system, Mirai aims to transform these systems into botnets that can launch substantial network-level and \ac{HTTP} flood attacks on servers. Mirai executes this by exploiting the default username and password combinations configured during the initialization of \ac{IoT} devices. The common expectation is that users will replace these default credentials. However, this often does not occur in practice, leading to devices remaining vulnerable to malicious intrusion. In such cases, hackers leverage scanning and bruteforce attacks to identify accessible devices to gain control over the device by injecting the Mirai malware. The Mirai attack follows the following sequence of events: (\textit{Scanning Stage}) The existing bots initiate a scan to identify potential new devices to infect. As the bots were deployed on two CM4 devices with limited processing power, the scanning process was significantly time-consuming. To expedite the bruteforce stage, we manually configured the target IP in the code; (\textit{Bruteforce Stage}) The bots then attempts to brute open Telnet ports on discovered devices utilizing a set of commonly used IoT device credentials. Upon successful bruteforce attempts, the bots report the pertinent device details and the successful credentials to the ScanListen server; (\textit{Loader Stage}) The \ac{CnC} server monitors the status of the ScanListen server and instructs the loader to inject a malicious binary onto the discovered device upon successful authentication. The Mirai malware was manually loaded onto the CM4 bots as they were found to be immune to Mirai infection; (\textit{Attack Stage}) The \ac{CnC} server then dispatches attack commands to the bots to initiate an attack on a specific victim \ac{IP}. 

We initiated eight vectors of the Mirai attack, specifically: ACK, \ac{DNS}, HTTP, GREETH, GREIP, SYN (SYN URG, SYN PUSH, SYN RST, SYN FIN, SYN-ACK), UDP, and UDP plain flood attacks. The UDP Plain flood attack is a simplified version of the \ac{UDP} flood, offering limited options but enabling a higher packet transmission rate. The GREETH and GREIP attacks inundate the target with malicious Generic Routing Encapsulation (GRE) encapsulated Ethernet and \ac{IP} packets, respectively. The GREETH assault includes Transparent Ethernet Bridging over GRE-encapsulated packets in its payload, while the GREIP attack encompasses solely \ac{IP} packets. Despite similar operational patterns, the GREETH attack incorporates an additional L2 frame. We altered various attack parameters in initiating the Mirai \ac{DDoS} assaults, some of which include the payload size, randomized source and destination ports, and type of service, as elaborated upon in Appendix Table~\ref{tab:mirai_params}. The resultant Mirai \ac{DDoS} attack data comprises two primary traffic types: \ac{CnC} traffic, capturing the interaction between the botmaster and the bots, as well as bot traffic, which represents the \ac{DDoS} attack activities. We also share the scanning and bruteforce traffic between the bots and the target device. 

\section{Network Traffic Feature Extraction and Importance Evaluation} \label{sec:feat_engineering}
This section is dedicated to exploring the procedures of feature extraction and importance evaluation in network traffic data. Our main interest lies in revealing inherent statistical tendencies and subtleties encapsulated in the network traffic data that have been generated. An overview of the data preprocessing stages, including the filtering of \ac{PCAP} files, is given in Section~\ref{sec:preprocessing}. We employ the CICFlowMeter tool for feature extraction and elucidate this process in Section~\ref{sec:feat_extraction}. In Section~\ref{sec:feat_imp}, we delve into feature importance analysis, providing an in-depth study of the most impactful features related to various types of network traffic.

\subsection{Data Filtering and Preprocessing}\label{sec:preprocessing}
 Following data capture, Wireshark was used to filter the obtained files, stored in the \ac{PCAP} format, based on the type of traffic each contained. Files containing malicious data underwent manual filtering to eliminate background traffic, which helped prevent contamination of the malicious files with benign data. The background traffic \ac{PCAP} helped determine what types of benign data packets the malicious traffic files needed to be filtered from. We noticed the rare presence of packets with random protocols in the files associated with \ac{DoS} attacks. As these are presumably part of the executed attack, they were not filtered out.

\subsection{Feature Extraction}\label{sec:feat_extraction}

While the primary objective of this study is not to contribute to the field of feature engineering, it is essential to describe the process we employed to extract valuable insights from our network traffic data. We utilized CICFlowMeter, a well-acknowledged tool frequently employed in intrusion detection literature. CICFlowMeter establishes a robust framework for extracting crucial features from traffic sessions. These sessions are defined based on bidirectional flows, a strategy consistent with the predominant network traffic object used for classification, compared to packets and unidirectional flows. Bidirectional flows offer a comprehensive network traffic perspective, facilitating precise and detailed examination. 
CICFlowMeter enables us to extract 75 distinct features from each bidirectional flow. The tool processes raw network traffic data maps the packets to their respective bidirectional flows, and then computes essential statistical features\footnote{For more details regarding the extracted features, please refer to the CICFlowMeter Github repository: \url{https://github.com/ahlashkari/CICFlowMeter}}. The processed data, represented in the form of these computed features, is provided in a structured \ac{CSV} file format. This format streamlines the subsequent stages of network traffic data analysis and interpretation. The \ac{CSV} files were labeled, incorporating three levels of classification such as ``Label" (Benign or Malicious), ``Traffic Type" (Audio, Background, Text, Video, Bruteforce, DoS, Information Gathering, Mirai), and ``Traffic Subtype" as listed in Table~\ref{tab:dataset_statistics}.

\begin{figure}[!t]
    \footnotesize
    \centering
    \begin{tabular}{c}
         \includegraphics[width=\columnwidth]{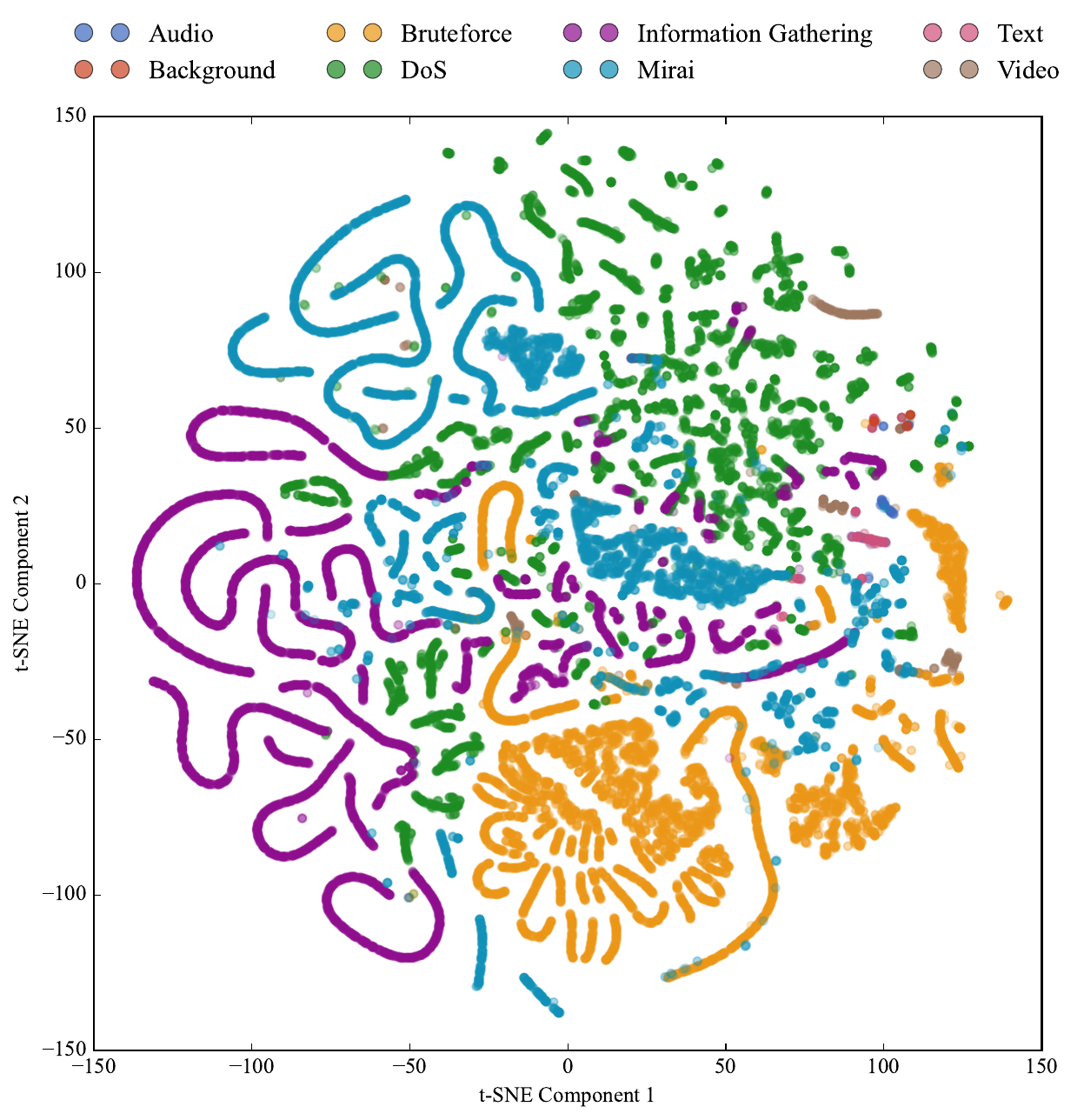}
    \end{tabular} 
    \caption{Clusters in network traffic data visualized using t-SNE.} 
    \label{fig:tsne}
\end{figure}  

To further understand the distribution and structure of our high-dimensional data, we employ \ac{t-SNE} for visualization. Figure~\ref{fig:tsne} presents the \ac{t-SNE} plot of our data, providing a clear visual summary of how our data points relate. From the plot, one can also discern the rich diversity inherent in the TII-SSRC-23 dataset. While distinct clusters corresponding to different traffic types are evident, the mingling of samples, especially within the malicious categories, underscores the multifaceted nature of intrusion patterns captured in our dataset. This intermingling, far from being a drawback, actually highlights the dataset's comprehensive coverage of a vast spectrum of attack vectors and behaviors.
  
\subsection{Feature Importance Analysis}\label{sec:feat_imp}
Before delving into the experimental phase of this study, it is critical to conduct a comprehensive analysis of feature importance. This analysis not only allows us to ascertain the relative significance of each feature and comprehend its bearing on the classification task but also provides insights for future work. Given the high dimensionality of our dataset, pinpointing the features that contribute most profoundly to our classification models' performance is vital. Additionally, this analysis is instrumental for future research that utilizes our shared dataset, as it provides valuable insights into model development within intrusion detection. This foundational understanding of feature importance could be leveraged to enhance the effectiveness of future intrusion detection models and strategies.  
 
We employed \ac{PFI} to compute the feature importance. \ac{PFI} works by randomly shuffling the values of one feature at a time and then evaluating the resultant effect on the model's performance. A marked decrease in the model's performance implies the shuffled feature's importance for the predictive task in question. However, evaluating feature importance should not entirely depend on a singular execution of \ac{PFI}. It is advisable to perform multiple runs per method and utilize various classifiers when assessing feature importance. This is because a feature's importance can fluctuate depending on the model's architecture and the specific run of the algorithm. We promote a more comprehensive understanding of feature importance by employing multiple methods and runs, providing a more robust foundation for our analysis. We employed three classifiers to calculate feature importance: the \ac{RF} classifier, the \ac{XGBoost} classifier, and the \ac{ET} classifier. These classifiers were selected by their efficiency and potential for parallelization, which permitted the experiment to be carried out within a feasible timeframe. It's also worth noting that, for each classifier, we conducted three separate runs of \ac{PFI}, thereby enhancing the reliability of our feature importance estimations.

\begin{figure}[!t]
    \footnotesize
    \centering
    \begin{tabular}{c}
         \includegraphics[width=\columnwidth]{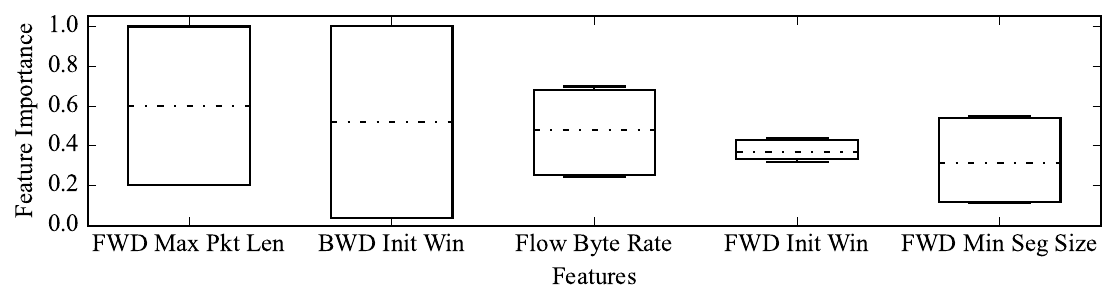} \\
         (a) Classifying network traffic as benign or malicious. \phantomsection \label{fig:feat_importance_a}\\
         \includegraphics[width=\columnwidth]{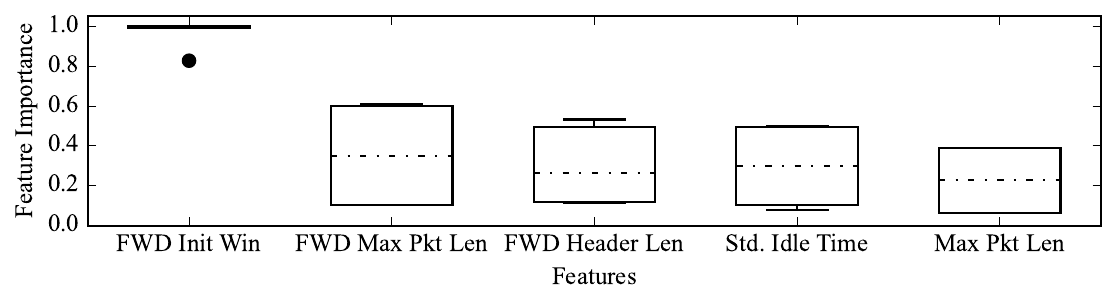} \\
         (b) Classifying network traffic given traffic types.
    \end{tabular} 
    \caption{Ranking of the five most critical features in network traffic classification. Plot (a) illustrates the five attributes distinguishing benign from malicious traffic. Plot (b) depicts the five principal features employed in segregating network traffic into various unique categories: audio, video, text, DoS, Mirai, and bruteforce attacks.} 
    \label{fig:feat_importance}
\end{figure}  

Two distinct feature importance experiments were conducted: (1) a binary classification experiment aimed at distinguishing benign from malicious traffic and (2) a multiclass classification experiment intended to identify specific types of network traffic. Boxplots of the feature importances for each scenario are presented in Figure~\ref{fig:feat_importance}. Plot (a) displays the top five features in distinguishing benign traffic from malicious ones. In contrast, plot (b) outlines the five most influential features in segregating network traffic into various unique categories, encompassing audio, video, text, \ac{DoS}, Mirai, and bruteforce attacks.

\begin{figure*}[!t]
    \centering
    \includegraphics[width=\textwidth]{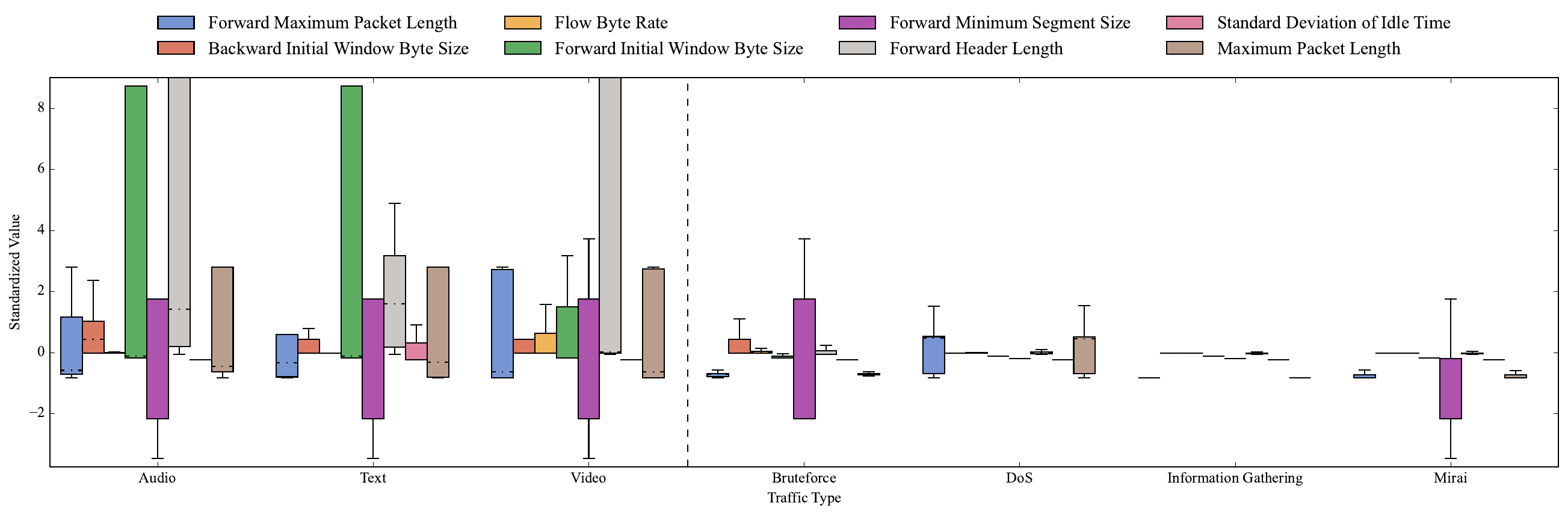}
    \caption{Variation of standardized feature values across traffic types.}
    \label{fig:feats_variation}
\end{figure*} 

Results from the feature importance experiment classifying benign vs. malicious traffic indicate that the top five most important attributes are Forward Maximum Packet Length (FWD Max Pkt Len), Backward Initial Window Byte Size (BWD Init Win), Flow Byte Rate (also referred as Flow Byte/s), Forward Initial Window Byte Size (FWD Init Win), and Forward Minimum Segment Size (FWD Min Seg Size). Notably, FWD Max Pkt Len and BWD Init Win present high feature importance scores, particularly in their third quartile values, implying a critical role in distinguishing benign and malicious network traffic. These features' broad range of importance values reflects their diverse influence across different classifiers and \ac{PFI} runs. Moreover, the Flow Byte Rate feature shows considerable variability in its importance, as evidenced by its interquartile range. Despite not reaching the upper limit seen in the first two features, it retains a notable importance score, making it a valuable contributor to traffic classification. In contrast, the FWD Init Win feature exhibits a relatively stable and moderate range of importance values, suggesting a steady but lesser contribution to network traffic classification. Finally, while not as impactful as the top-ranking features, the FWD Min Seg Size feature still contributes to the classification task. Its median importance score, though lower, provides a meaningful addition to the overall classification task. 

Results from the feature importance experiment, aimed at classifying network traffic into various unique categories, indicate that the top five most important attributes are Forward Initial Window Byte Size (FWD Init Win), Forward Maximum Packet Length (FWD Max Pkt Len), Forward Header Length (FWD Header Len), Standard Deviation of Idle Time (Std. Idle Time), and Maximum Packet Length (Max Pkt Len). The FWD Init Win feature is the most significant, supported by its nearly maximal feature importance scores across the first, second, and third quartiles. Its consistently high importance demonstrated across multiple classifiers and \ac{PFI} runs, underscores its pivotal role in differentiating between various types of network traffic. Remarkably, FWD Init Win is one of the top five important features in both experiments, attesting to its relevance across distinct classification tasks. The other four features also contribute significantly to the classification task, with varying importance scores. FWD Max Pkt Len, particularly in its third quartile, substantially influences traffic classification. Additionally, FWD Header Len and Std. Idle Time plays important roles, enhancing the model's ability to distinguish between traffic types. Max Pkt Len, although not scoring as high as the others, still contributes notably to the overall classification task. These top five attributes, especially FWD Init Win featured in both experiments, play a vital role in effectively classifying network traffic.

Given the most important features identified from the feature importance analysis, we can now examine their raw values across different traffic types. Figure~\ref{fig:feats_variation} presents the standardized feature values for the top eight most important features, allowing us to identify significant variations among the traffic types. Across the Video, Audio, and Text traffic types, we notice a notable variation in the values of the features compared to the \ac{DoS}, Mirai, and Bruteforce traffic types. There seems to be a consistent pattern for the first three traffic types, where the feature values exhibit a more widespread distribution, covering a larger range of values. In contrast, the \ac{DoS}, Mirai, and Bruteforce traffic types show a more concentrated distribution of feature values, with relatively lesser variations. Moreover, we can identify several features with distinct characteristics among the first three traffic types. For instance, FWD Max Pkt Len stands out with relatively high variability in the values across the Video, Audio, and Text traffic types. In contrast, features like FWD Init Win and FWD Header Len exhibit relatively stable and consistent values across the benign traffic types. We notice a different trend when examining the malicious traffic types (\ac{DoS}, Mirai, and Bruteforce). The features display more uniform values, indicating less variability across these traffic types. Features such as FWD Min Seg Size and FWD Header Len show particularly distinct characteristics compared to the benign traffic types, reinforcing their relevance in distinguishing between benign and malicious traffic. 

\section{Experimental Evaluation and Baseline Results}\label{sec:experiments}
In this section, we evaluate supervised and unsupervised methodologies to establish firm baseline performances for intrusion detection utilizing our dataset. This undertaking serves two functions. Firstly, it equips future research that leverages our dataset with crucial insights and performance benchmarks. Secondly, it offers robust baselines for two essential tasks in network security: supervised intrusion detection and unsupervised intrusion detection via \ac{OOD} detection, that is, network anomaly detection. Through this, we enable the comparison of emerging models and methodologies using our shared dataset, thereby promoting the development of more effective intrusion detection systems. Section~\ref{sec:supervised} details the application of supervised methodologies to distinguish various types of network traffic while simultaneously acknowledging the inherent limitations of these methods when dealing with unseen attacks absent from the training data. In contrast, Section~\ref{sec:unsupervised} investigates the use of unsupervised approaches for anomaly detection, emphasizing the need to incorporate a wide variety of real-world traffic patterns to boost model robustness and adaptability to changing traffic distributions. 

\subsection{Data Handling and Experimental Design}\label{sec:setup}
The preprocessing phase involved removing unnecessary columns and duplicates. The columns removed were source IP and port, destination IP and port, and flow identifier, allowing us to focus on the most pertinent features for our analysis. We applied normalization and standard scaling techniques to address disparities in the scales of different features. Missing data was handled using two different strategies based on the nature of the data. Missing values in numerical data were substituted with the mean value of the respective feature.
In contrast, missing values were replaced with the most frequent category for categorical data. One-hot encoding was employed specifically for the `protocol' feature, the only categorical variable in our dataset. We refrained from performing any form of dimensionality reduction. We experimented with balancing the dataset using the \ac{SMOTE}. However, this did not result in any significant performance improvement.

To evaluate the models, we employed several metrics, including the F1 score, \ac{AUROC}, and \ac{AUC-PR}. The F1 score balances precision and recall and provides an overall assessment of a model's accuracy. The F1 score we employed uses the macro average, the unweighted mean of the F1 scores for each class. The \ac{AUROC} measures a model's capability to distinguish between classes, with a higher \ac{AUROC} indicating better performance. The \ac{AUC-PR} summarizes the precision-recall curve and is particularly useful in scenarios with class imbalances. These metrics were chosen based on our problem's characteristics and the need to assess the models from various perspectives.   

\subsection{Baselines for Supervised-based Intrusion Detection} \label{sec:supervised}
Our experiments for the supervised classification are carried out in three steps: (1) a binary classification to differentiate between benign and malicious traffic, (2) a multiclass classification to categorize diverse types of traffic, and (3) a multiclass classification to classify the traffic into subtypes further.
In our supervised experiment, we opted for the following classifiers: \ac{RF}, \ac{DT}, \ac{ET}, \ac{MLP}, \ac{SVM}, and \ac{XGBoost}. Although K-Nearest Neighbors was initially considered, it was later omitted from our selection due to its below-average experiment results. These models were chosen due to their widespread utilization, interpretability, and robustness in dealing with various classification problems.  
 
\begin{table}[t]
    \centering
    \caption{Baseline results (\%) of ML models on our published dataset for supervised network intrusion detection tasks. These results provide a baseline for future research and comparison with emerging models and methodologies.}  
    \begin{tabularx}{\columnwidth}{lCCCC} 
        \toprule
            \textbf{Models}        & \textbf{Accuracy} & \textbf{F1 Score}  &  \textbf{\ac{AUROC}}   &  \textbf{\ac{AUC-PR}} \\
        \midrule
        \multicolumn{5}{c}{\textbf{Benign vs. Malicious -- Binary Classification Results}} \\
        \midrule
            SVM           &           99.84 &           57.87 &         97.61 &  \textbf{100} \\
            MLP           &           99.99 &           89.48 &         99.83 &   \textbf{100} \\
            Decision Tree &    \textbf{100} &           96.87 &         97.24 &   \textbf{100} \\
            Random Forest &    \textbf{100} &           98.01 &         98.62 &   \textbf{100} \\
            Extra Trees   &    \textbf{100} &           98.60 &         98.62 &   \textbf{100} \\
            XGBoost       &    \textbf{100} &  \textbf{98.79} &  \textbf{100} &   \textbf{100} \\
        \midrule
        \multicolumn{5}{c}{\textbf{Network Traffic Types -- Multiclass Classification Results}} \\
        \midrule
            SVM           &  97.73 &  61.66 &  96.45 &   72.44 \\
            MLP           &  99.94 &  75.60 &  97.81 &   82.62 \\
            Decision Tree &  99.98 &  94.84 &  97.12 &   93.21 \\
            Extra Trees   &  99.98 &  96.71 &  99.49 &   97.46 \\
            Random Forest &  99.98 &  97.28 &  99.53 &   97.66 \\
            XGBoost       & \textbf{ 99.99} &  \textbf{97.31} &  \textbf{99.80} &   \textbf{98.34} \\
        \midrule
        \multicolumn{5}{c}{\textbf{Network Traffic Subtypes -- Multiclass Classification Results}} \\
        \midrule
            MLP           &     99.71 &  78.41 &  99.07 &   85.63 \\
            SVM           &     99.29 &  80.57 &  97.39 &   81.80 \\
            Decision Tree &     99.74 &  90.81 &  96.33 &   90.11 \\
            XGBoost       &     \textbf{99.79} &  92.73 &  \textbf{99.77} &   \textbf{94.45} \\
            Random Forest &     99.75 &  93.05 &  98.61 &   92.93 \\ 
            Extra Trees   &     99.76 &  \textbf{93.36} &  98.77 &   92.95 \\
        \bottomrule        
    \end{tabularx} 
    \label{tab:sup_exp}
\end{table}    

Each classifier underwent a hyperparameter tuning process using grid search. The grid search resulted in the following approximated optimal hyperparameters. For \ac{RF}, the maximum tree depth was found to be `none', the minimum number of samples required to split a node was 2, and the number of estimators used was 100. For \ac{DT}, the function for measuring the quality of splits was `entropy', the maximum tree depth was `none', the minimum number of samples required at a leaf node was 1, and the minimum number of samples required to split an internal node was 5. For \ac{ET}, the function for measuring the quality of splits was `entropy', the maximum tree depth was `none', the minimum number of samples required to split a node was 4, and the number of estimators used was 200. For the \ac{MLP}, the activation function was `tanh', the L2 penalty (regularization term) parameter was 0.0001, the configuration for the number of neurons in the hidden layers was (64, 64), and the solver for weight optimization was `adam'. For \ac{XGBoost}, the maximum depth of the trees was 6, the learning rate was 0.1, the subsample ratio of the training instances was 1, the number of gradient-boosted trees was 200, and the subsample ratio of columns for each split, in each level, was 0.5. Finally, for the \ac{SVM}, the penalty parameter of the error term was 1, the kernel coefficient was `scale', and the function used in the algorithm was `linear'.

Table~\ref{tab:sup_exp} presents the mean performance metrics obtained from three separate runs of each method from three separate runs of each model. Binary classification results showed high performance from all models for distinguishing benign and malicious traffic, with \ac{SVM} having the lowest F1 score of 57.87 (accuracy 99.84) and \ac{XGBoost} having the highest F1 score of 98.79 (\ac{AUROC} 100). Multiclass classification for traffic types saw similar performance, with \ac{SVM} lowest and \ac{XGBoost} highest (F1 score 97.31, \ac{AUROC} 99.80). \ac{MLP}, \ac{DT}, \ac{ET}, and \ac{RF} exceeded 99.94 accuracies. Traffic subtype results followed this trend, with \ac{MLP} and \ac{SVM} lagging (F1 scores of 78.41 and 80.57, respectively) and \ac{ET} leading (F1 score of 93.36).

The results demonstrate that the selected classifiers generally performed well in our dataset's binary and multiclass classifications. However, the performance was not uniform across all models in binary tasks, with \ac{SVM} and \ac{MLP} classifiers yielding less satisfactory F1 scores. Conversely, the \ac{XGBoost} and \ac{ET} classifiers excelled in all experiments, proficiently classifying benign and malicious traffic and differentiating various traffic types and subtypes. As we subdivided network traffic into more refined categories, a noticeable decline in the performance of our methods became apparent, underscoring the increased challenge in finer-grained classifications. For a detailed understanding of the performance, refer to the classification results for each class in each experiment, provided in the Appendix (Tables \ref{tab:exp_sup_binary_detailed}, \ref{tab:exp_sup_multiclass_type_detailed}, and \ref{tab:exp_sup_multiclass_subtype_detailed}). Table~\ref{tab:exp_sup_binary_detailed} provides the XGBoost precision, recall, and F1 score for benign and malicious traffic. Table~\ref{tab:exp_sup_multiclass_type_detailed} offers details on the XGBoost precision, recall, and F1 score for each traffic \textit{type}, while Table~\ref{tab:exp_sup_multiclass_subtype_detailed} delineates the Extra Trees precision, recall, and F1 score for each traffic \textit{subtype}. This additional information enhances our understanding of the models' effectiveness across diverse traffic types and subtypes.

\subsection{Baselines for Anomaly-based Intrusion Detection} \label{sec:unsupervised}
We formulate anomaly-based intrusion detection as an unsupervised task, conceptualizing it as an \ac{OOD} detection problem. In this configuration, the in-distribution is represented by normal data, the only data type used during model training. During testing, both normal and malicious traffic are introduced, the distributions of which should ideally be separable. For this experiment, the focal evaluation metrics are the \ac{AUROC} and the F1 score, computed at both the 99th percentile and maximum threshold of the scores obtained from the training set. The maximum threshold, a well-known thresholding technique, is particularly effective when the normal and malicious traffic score distributions do not overlap, thereby representing a distinct separation between these classes. Conversely, the 99th percentile threshold is employed to handle situations with extreme maximum normal scores.

The anomaly detection methods selected for our unsupervised experiments include: \ac{IF}, \ac{KDE}, \ac{LOF}, \ac{OC-SVM}, and \ac{Deep SVDD}~\cite{ruff2018deep}. We conducted a grid search for each method to tune hyperparameters. The approximate optimal hyperparameters derived from this procedure are: For \ac{OC-SVM}, we set the kernel function to 'linear', $\gamma$ to 'auto', and $\nu$ to 0.1. We set the kernel function to Gaussian for \ac{KDE} and the bandwidth to 'auto'. For \ac{IF}, we set the number of estimators to 2000. For \ac{LOF}, we set the number of neighbors to 20 and the leaf size to 30. The \ac{Deep SVDD} was implemented using Pytorch 2.0.1, employing an encoder and decoder \ac{MLP} architecture for pre-training while minimizing the mean squared error over 1000 epochs. This approach facilitates the encoder in learning the nuances of a normal distribution. After the preliminary phase of encoder pre-training, the center point is calculated, and the decoder component is eliminated. Subsequently, the encoder undergoes further training to reduce the distance between projected embeddings and the center point. The underlying logic is that by adopting this strategy, the model will be adept at mapping normal samples closer to the central point while unable to do so as efficiently for the samples not included during the training process. The deviation between the projected and center embedding is used as a scoring metric during testing. The pre-training and training phases employed Adam optimization, with a learning rate 1e-4 and an L2 penalty of 1e-6. The encoder and decoder consist of a 79-neuron layer with ReLU activation, followed by a linear layer. The latent space has been dimensioned to 20.

\begin{table}[t]
    \centering
    \caption{Baseline results (\%) for anomaly-based intrusion detection methods. Metrics are presented for two different threshold settings: the 99th percentile and the maximum value. The table compares each model's \ac{AUROC}, precision, recall, and F1 score under each threshold setting.}\label{tab:exp_unsup}
    \begin{tabularx}{\columnwidth}{lCCCCCCC} 
        \toprule
        \multirow{2}{*}{\textbf{Models}} & \multirow{2}{*}{\textbf{AUC}} & \multicolumn{3}{c}{\textbf{99th Threshold}} & \multicolumn{3}{c}{\textbf{Maximum Threshold}} \\
        \cmidrule(lr){3-5} \cmidrule(lr){6-8}
        & & \textbf{Prec.} & \textbf{Rec.} & \textbf{F1} & \textbf{Prec.} & \textbf{Rec.} & \textbf{F1} \\
        \midrule
        IF & 58.21 & 38.46 & 0.79 & 1.54 & 0.0 & 0.0 & 0.0 \\ 
        KDE & 64.19 & 64.95 & 95.43 & 77.3 & 64.95 & 95.43 & 77.3 \\  
        LOF & 92.35 & 42.11 & 1.26 & 2.45 & 100.0 & 0.31 & 0.63 \\  
        OC-SVM & 96.64 & 99.98 & 57.99 & 73.41 & 99.98 & 9.16 & 16.79 \\ 
        Deep SVDD & \textbf{97.84} & 99.98 & 99.68 & \textbf{99.83} & 99.98 & 99.54 & \textbf{99.76} \\ 
        \bottomrule
    \end{tabularx} 
\end{table} 

Table~\ref{tab:exp_unsup} presents the mean performance metrics for each anomaly-based intrusion detection method analyzed, obtained from three separate runs of each model, all with distinct seed settings. 
The results indicate that the models have significant variations in their performance. For instance, the \ac{IF} model struggled to distinguish between normal and anomalous traffic, resulting in the lowest \ac{AUROC} of 58.21. This performance equated to a modest F1 score of 1.54 at the 99th percentile threshold. The model could not identify anomalies at the maximum threshold, yielding an F1 score of 0.0. KDE exhibited a satisfactory performance, registering an AUROC of 64.19. With an F1 score of 77.3 at both the 99th percentile and maximum thresholds, the KDE model demonstrated consistency across the two threshold settings. Following the KDE, \ac{Deep SVDD} showed exceptional performance, registering the highest \ac{AUROC} of 97.84 and a notable F1 score of 99.83 at the 99th percentile threshold. \ac{Deep SVDD} maintained a high F1 score of 99.76 even at the highest threshold, highlighting its stable performance across both threshold settings. The performance of \ac{LOF}, and \ac{OC-SVM} models was inconsistent. Interestingly, the \ac{OC-SVM} model showed high precision but observed a notable decrease in the recall and, therefore, the F1 score at the maximum threshold.

The outcomes highlight the variation in model performance and emphasize the significant effect of threshold selection on said performance. This highlights the necessity for meticulous threshold selection when evaluating unsupervised anomaly detection methods. Given the intricate nature of network anomaly detection, more sophisticated strategies are commonly needed for effective anomaly identification. Take, for example, ARCADE~\cite{lunardi2022arcade}, which implements a \ac{DL} strategy, leveraging an adversarially regularized 1D-convolutional neural network autoencoder to learn the normal traffic pattern from raw network data. Our dataset, including raw traffic, aligns well with these advanced techniques. The strong performance of \ac{Deep SVDD} in network traffic analysis further reinforces the value of adopting these advanced techniques. 

\section{Conclusion} \label{sec:conclusion}
Addressing the widespread challenge in public network traffic datasets where there is an overrepresentation of benign and a scarcity of diverse malicious network traffic, we introduce the TII-SSRC-23 dataset. We emphasize the importance of data diversity in enhancing \ac{IDS} efficacy within \ac{ML}-based paradigms. TII-SSRC-23 dataset encompasses a wide spectrum of benign and malicious traffic patterns, including 32 benign and malicious traffic subtypes with 26 unique attacks launched, each enriched with many variations in traffic parameters.  
Although the imbalance towards malicious samples of our dataset may appear to be a drawback, we highlight that this reflects the diversity present in the malicious traffic. As previously mentioned, the representation of benign examples can be enriched with traffic from the aforementioned public datasets.
By exploring feature importance analysis, we have successfully unearthed the generated data's inherent statistical tendencies and intricacies. Moreover, our experimental evaluations established benchmark performance for each subtype. These benchmarks not only serve as a baseline for upcoming research but also underscore the importance of using both supervised and unsupervised methodologies in ensuring comprehensive security coverage against a wide array of network threats.

Future improvements upon our research could benefit from expanding the TII-SSRC-23 dataset by merging it with other benign datasets, amplifying the diversity of benign traffic types, and enhancing the dataset's representativeness. Furthermore, the performance of \ac{IDS} models trained on our data could be rigorously tested in real-world deployment scenarios to assess their effectiveness under actual operating conditions. The insights from this paper can steer future research towards prioritizing traffic diversity to capture the complexities of network traffic, thereby strengthening the development of intrusion detection systems to address evolving network security challenges effectively. 
  
{
    \small
    \bibliographystyle{IEEEtranN}
    \bibliography{main} 

\begin{thebibliography}{24}
\providecommand{\natexlab}[1]{#1}
\providecommand{\url}[1]{#1}
\csname url@samestyle\endcsname
\providecommand{\newblock}{\relax}
\providecommand{\bibinfo}[2]{#2}
\providecommand{\BIBentrySTDinterwordspacing}{\spaceskip=0pt\relax}
\providecommand{\BIBentryALTinterwordstretchfactor}{4}
\providecommand{\BIBentryALTinterwordspacing}{\spaceskip=\fontdimen2\font plus
\BIBentryALTinterwordstretchfactor\fontdimen3\font minus
  \fontdimen4\font\relax}
\providecommand{\BIBforeignlanguage}[2]{{%
\expandafter\ifx\csname l@#1\endcsname\relax
\typeout{** WARNING: IEEEtranN.bst: No hyphenation pattern has been}%
\typeout{** loaded for the language `#1'. Using the pattern for}%
\typeout{** the default language instead.}%
\else
\language=\csname l@#1\endcsname
\fi
#2}}
\providecommand{\BIBdecl}{\relax}
\BIBdecl

\bibitem[Draper-Gil et~al.(2016)Draper-Gil, Lashkari, Mamun, and
  Ghorbani]{draper2016CICFlow}
G.~Draper-Gil, A.~H. Lashkari, M.~S.~I. Mamun, and A.~A. Ghorbani,
  ``Characterization of encrypted and vpn traffic using time-related,'' in
  \emph{Proceedings of the 2nd international conference on information systems
  security and privacy (ICISSP)}, 2016, pp. 407--414.

\bibitem[{Massachusetts Institute of Technology}(1998)]{mit1998DARPA}
{Massachusetts Institute of Technology}, ``1998 darpa intrusion detection
  evaluation dataset,''
  \url{https://www.ll.mit.edu/r-d/datasets/1998-darpa-intrusion-detection-evaluation-dataset},
  1998.

\bibitem[Stolfo()]{stolfoKDD}
S.~Stolfo, ``Kdd cup 1999 dataset,'' Date last accessed 22-June-2018. [link].
  URL \url{http://kdd.ics.uci.edu/databases/kddcup99/kddcup99.html}.

\bibitem[Tavallaee et~al.(2009)Tavallaee, Bagheri, Lu, and
  Ghorbani]{tavallaee2009NSLKDD}
M.~Tavallaee, E.~Bagheri, W.~Lu, and A.~A. Ghorbani, ``A detailed analysis of
  the kdd cup 99 data set,'' in \emph{2009 IEEE symposium on computational
  intelligence for security and defense applications}.\hskip 1em plus 0.5em
  minus 0.4em\relax Ieee, 2009, pp. 1--6.

\bibitem[Song et~al.(2011)Song, Takakura, Okabe, Eto, Inoue, and Nakao]{kyoto}
J.~Song, H.~Takakura, Y.~Okabe, M.~Eto, D.~Inoue, and K.~Nakao, ``Statistical
  analysis of honeypot data and building of kyoto 2006+ dataset for nids
  evaluation,'' pp. 29--36, 2011.

\bibitem[Shiravi et~al.(2012)Shiravi, Shiravi, Tavallaee, and
  Ghorbani]{shiravi12iscx}
A.~Shiravi, H.~Shiravi, M.~Tavallaee, and A.~A. Ghorbani, ``Toward developing a
  systematic approach to generate benchmark datasets for intrusion detection,''
  \emph{Computers \& Security}, vol.~31, no.~3, pp. 357--374, 2012.

\bibitem[Sharafaldin et~al.(2018)Sharafaldin, Lashkari, and
  Ghorbani]{sharafaldin2018CICIDS}
I.~Sharafaldin, A.~H. Lashkari, and A.~A. Ghorbani, ``Toward generating a new
  intrusion detection dataset and intrusion traffic characterization.''
  \emph{ICISSp}, vol.~1, pp. 108--116, 2018.

\bibitem[Moustafa and Slay(2015)]{moustafa15UNSW}
N.~Moustafa and J.~Slay, ``{UNSW-NB15}: A comprehensive data set for network
  intrusion detection systems,'' in \emph{Military Communications and
  Information Systems Conference (MilCIS)}.\hskip 1em plus 0.5em minus
  0.4em\relax IEEE, 2015, pp. 1--6.

\bibitem[Gringoli et~al.(2009)Gringoli, Salgarelli, Dusi, Cascarano, Risso, and
  Claffy]{gringoli2009UNIBS}
F.~Gringoli, L.~Salgarelli, M.~Dusi, N.~Cascarano, F.~Risso, and K.~Claffy,
  ``Gt: picking up the truth from the ground for internet traffic,'' \emph{ACM
  SIGCOMM Computer Communication Review}, vol.~39, no.~5, pp. 12--18, 2009.

\bibitem[Garcia et~al.(2014)Garcia, Grill, Stiborek, and Zunino]{garcia2014CTU}
S.~Garcia, M.~Grill, H.~Stiborek, and A.~Zunino, ``An empirical comparison of
  botnet detection methods,'' \emph{Computers \& Security}, vol.~45, pp.
  100--123, 2014.

\bibitem[Bhuyan et~al.(2015)Bhuyan, Bhattacharyya, and Kalita]{bhuyan2015TUIDS}
M.~H. Bhuyan, D.~K. Bhattacharyya, and J.~K. Kalita, ``Towards generating
  real-life datasets for network intrusion detection.'' \emph{Int. J. Netw.
  Secur.}, vol.~17, no.~6, pp. 683--701, 2015.

\bibitem[Alkasassbeh et~al.(2016)Alkasassbeh, Al-Naymat, Hassanat, and
  Almseidin]{alkasassbeh2016DDoS}
M.~Alkasassbeh, G.~Al-Naymat, A.~B. Hassanat, and M.~Almseidin, ``Detecting
  distributed denial of service attacks using data mining techniques,''
  \emph{International Journal of Advanced Computer Science and Applications},
  vol.~7, no.~1, 2016.

\bibitem[Jazi et~al.(2017)Jazi, Gonzalez, Stakhanova, and
  Ghorbani]{jazi2017CICDoS}
H.~H. Jazi, H.~Gonzalez, N.~Stakhanova, and A.~A. Ghorbani, ``Detecting
  http-based application layer dos attacks on web servers in the presence of
  sampling,'' \emph{Computer Networks}, vol. 121, pp. 25--36, 2017.

\bibitem[Moustafa(2019)]{moustafaBotIoT}
\BIBentryALTinterwordspacing
N.~Moustafa, ``The bot-iot dataset,'' 2019. [Online]. Available:
  \url{https://dx.doi.org/10.21227/r7v2-x988}
\BIBentrySTDinterwordspacing

\bibitem[Almaraz-Rivera et~al.(2022)Almaraz-Rivera, Perez-Diaz,
  Cantoral-Ceballos, Botero, and Trejo]{almaraz2022LATAM}
J.~G. Almaraz-Rivera, J.~A. Perez-Diaz, J.~A. Cantoral-Ceballos, J.~F. Botero,
  and L.~A. Trejo, ``Toward the protection of iot networks: Introducing the
  latam-ddos-iot dataset,'' \emph{IEEE Access}, vol.~10, pp.
  106\,909--106\,920, 2022.

\bibitem[Dadkhah et~al.(2022)Dadkhah, Mahdikhani, Danso, Zohourian, Truong, and
  Ghorbani]{dadkhah2022CICIoT}
S.~Dadkhah, H.~Mahdikhani, P.~K. Danso, A.~Zohourian, K.~A. Truong, and A.~A.
  Ghorbani, ``Towards the development of a realistic multidimensional iot
  profiling dataset,'' in \emph{2022 19th Annual International Conference on
  Privacy, Security \& Trust (PST)}.\hskip 1em plus 0.5em minus 0.4em\relax
  IEEE, 2022, pp. 1--11.

\bibitem[Ferrag et~al.(2022)Ferrag, Friha, Hamouda, Maglaras, and
  Janicke]{ferrag2022EdgeIoT}
M.~A. Ferrag, O.~Friha, D.~Hamouda, L.~Maglaras, and H.~Janicke,
  ``Edge-iiotset: A new comprehensive realistic cyber security dataset of iot
  and iiot applications for centralized and federated learning,'' \emph{IEEE
  Access}, vol.~10, pp. 40\,281--40\,306, 2022.

\bibitem[Moustafa(2021)]{moustafa2021TONIoT}
N.~Moustafa, ``A new distributed architecture for evaluating ai-based security
  systems at the edge: Network ton\_iot datasets,'' \emph{Sustainable Cities
  and Society}, vol.~72, p. 102994, 2021.

\bibitem[Meidan et~al.(2018)Meidan, Bohadana, Mathov, Mirsky, Shabtai,
  Breitenbacher, and Elovici]{meidan2018Nbaiot}
Y.~Meidan, M.~Bohadana, Y.~Mathov, Y.~Mirsky, A.~Shabtai, D.~Breitenbacher, and
  Y.~Elovici, ``N-baiot—network-based detection of iot botnet attacks using
  deep autoencoders,'' \emph{IEEE Pervasive Computing}, vol.~17, no.~3, pp.
  12--22, 2018.

\bibitem[Sommer and Paxson(2010)]{sommerMLIDS}
R.~Sommer and V.~Paxson, ``Outside the closed world: On using machine learning
  for network intrusion detection,'' in \emph{2010 IEEE Symposium on Security
  and Privacy}, 2010, pp. 305--316.

\bibitem[Maci{\'a}-Fern{\'a}ndez et~al.(2010)Maci{\'a}-Fern{\'a}ndez,
  Rodr{\'\i}guez-G{\'o}mez, and D{\'\i}az-Verdejo]{macia2010defense}
G.~Maci{\'a}-Fern{\'a}ndez, R.~A. Rodr{\'\i}guez-G{\'o}mez, and J.~E.
  D{\'\i}az-Verdejo, ``Defense techniques for low-rate dos attacks against
  application servers,'' \emph{Computer Networks}, vol.~54, no.~15, pp.
  2711--2727, 2010.

\bibitem[Antonakakis et~al.(2017)Antonakakis, April, Bailey, Bernhard,
  Bursztein, Cochran, Durumeric, Halderman, Invernizzi, Kallitsis,
  et~al.]{mirai}
M.~Antonakakis, T.~April, M.~Bailey, M.~Bernhard, E.~Bursztein, J.~Cochran,
  Z.~Durumeric, J.~A. Halderman, L.~Invernizzi, M.~Kallitsis \emph{et~al.},
  ``Understanding the mirai botnet,'' in \emph{26th USENIX security symposium
  (USENIX Security 17)}, 2017, pp. 1093--1110.

\bibitem[Ruff et~al.(2018)Ruff, Vandermeulen, Goernitz, Deecke, Siddiqui,
  Binder, M{\"u}ller, and Kloft]{ruff2018deep}
L.~Ruff, R.~Vandermeulen, N.~Goernitz, L.~Deecke, S.~A. Siddiqui, A.~Binder,
  E.~M{\"u}ller, and M.~Kloft, ``Deep one-class classification,'' in
  \emph{International conference on machine learning}.\hskip 1em plus 0.5em
  minus 0.4em\relax PMLR, 2018, pp. 4393--4402.

\bibitem[Lunardi et~al.(2022)Lunardi, Lopez, and Giacalone]{lunardi2022arcade}
W.~T. Lunardi, M.~A. Lopez, and J.-P. Giacalone, ``{ARCADE}: Adversarially
  {R}egularized {C}onvolutional {A}utoencoder for {N}etwork {A}nomaly
  {D}etection,'' \emph{IEEE Transactions on Network and Service Management},
  2022.

\end{thebibliography}
}


\appendices
\section{Problem Notation}\label{sec:prob_notation}

Let us formalize the concepts of unidirectional and bidirectional network flows. Consider a network where packets are transmitted between different endpoints. A packet $p$ can be defined as a tuple $p = (s_{\text{ip}}, s_{\text{prt}}, d_{\text{ip}}, d_{\text{prt}}, \tau)$, where $s_{\text{ip}}$ is the source IP address, $s_{\text{prt}}$ is the source port, $d_{\text{ip}}$ is the destination IP address, $d_{\text{prt}}$ is the destination port, and $\tau$ is the transport-level protocol used. The arrival of each packet is indicated by its corresponding timestamp $t$.

\subsection{Unidirectional Network Flow}
A unidirectional network flow $\mathcal{F} = (p_1, p_2, \ldots, p_n)$, commonly referred to as network flow, represents a sequence of $n$ packets that share the same 5-tuple, i.e., for any pair of packets $p_i$ and $p_j$, where $i, j \in \{1, 2, \ldots, n\}$, we have that $p_i = p_j$. Additionally, the packets within the flow are ordered based on their arrival timestamps, such that $t_{i} < t_{i + 1}$, where $i \in {1,2,\ldots,n-1}$. Here, $p_i$ represents the $i$-th packet in the sequence, and $t_i$ represents the timestamp of the $i$-th packet. 

\begin{table}[t]
    \centering
    \caption{IDS Datasets Malicious Traffic}
    \label{tab:dataset_attacks}
    \begin{tabularx}{\columnwidth}{@{}>{\raggedright\arraybackslash}p{2.5cm}X}
        \toprule
        \textbf{Dataset} & \textbf{Attacks} \\
        \midrule
        DARPA98 & DoS, privilege escalation (R2L, U2R), probing \\
        \midrule
        KDD99 & DoS, privilege escalation (R2L, U2R), probing \\
        \midrule
        NSL-KDD & DoS, privilege escalation (R2L, U2R), probing \\
        \midrule
        Kyoto 2006+ & DoS, backscatter, malware, port scans, shellcode, exploits \\
        \midrule
        UNIBS & None \\
        \midrule
        TUIDS & Botnet, DoS, IRC botnet DDoS, probing, coordinated port scan, U2R using bruteforce SSH \\
        \midrule
        ISCX 2012 & Infiltrating, HTTP DoS, IRC Botnet DDoS, SSH Bruteforce \\
        \midrule
        CTU-13 & Botnets (Menti, Murlo, Neris, NSIS, Rbot, Sogou, Virut) \\
        \midrule
        UNSW-NB15 & Backdoors, DoS, exploits, fuzzers, generic, port scans, reconnaissance, shellcode, spam, worms \\
        \midrule
        DDoS 2016 & DDoS (HTTP, SIDDoS, Smurf \ac{ICMP}, UDP) \\
        \midrule
        CICIDS 2017 & Botnet (Ares), DoS/DDoS, XSS, heartbleed, infiltration, SSH bruteforce, SQL injection \\
        \midrule
        CIC DoS & Application layer DoS attacks (high- and low-volume HTTP DoS) \\
        \midrule
        BoT-IoT & Probing (port scan, OS fingerprinting), DoS/DDoS (HTTP, TCP, UDP), information theft (data theft, keylogging) \\
        \midrule
        LATAM-DDoS-IoT & DoS and DDoS attacks (HTTP, TCP, UDP) \\
        \midrule
        CIC IoT & DoS (HTTP, TCP, UDP), RTSP Bruteforce Attack \\
        \midrule
        Edge-IIoTset & DoS/DDoS (HTTP, ICMP, TCP SYN, UDP), Information Gathering (Port scanning, OS fingerprinting, Vulnerability scanning), MitM (\ac{DNS} and ARP spoofing), Injection attacks (XSS, SQL injection, uploading attack), Malware (backdoor, password cracking, ransomware)\\
        \midrule
        N-Baiot & Botnets (Mirai, BASHLITE) \\
        \midrule
        TON-IoT & Scanning, DoS, DDoS, ransomware, backdoor, injection, XSS, password cracking, MitM \\
        \midrule
        TII-SSRC-23 & DoS (HTTP, ICMP, MAC, UDP, TCP SYN, TCP ACK, TCP PSH, TCP RST, TCP FIN, TCP URG, TCP ECN, TCP CWR), Information Gathering (TCP Port, UDP Port, Ping, OS, Version, Script scans), Bruteforce (DNS, FTP, HTTP, Telnet, SSH), Botnet (Mirai) \\
        \bottomrule
    \end{tabularx}
\end{table}
 
\begin{table*}[t]
\centering
\caption{Detailed overview of the tools, parameters, and combinations employed for the generation of benign traffic.}\label{tab:ben_params}
\begin{tabularx}{\textwidth}{@{}lXX>{\hsize=7cm}X>{\hsize=2.5cm}X@{}}
\toprule
\textbf{Traffic Type} & \textbf{Traffic Subtype} & \textbf{Tool} & \textbf{Parameters Varied} & \textbf{Combinations} \\
\midrule
Audio & Audio & Mumble & Audio message length\par{}  5\% disconnection rate \par{} Network interference: low, mid, high & 1 \\
\midrule
Background & Background & -- &  Network interference: low, mid, high & 1 \\
\midrule
Text & Text & Mumble &  Text message length\par{}  5\% disconnection rate \par{} Network interference: low, mid, high & 1 \\
\midrule
Video & HTTP\par{} RTP/TS \par{} UDP  & VLC & Video resolution: 240p, 360p, 480p, 720p, 1080p\par{}  Audio bitrate: 96 to 192\par{}  Video bitrate: 800 to 3500\par{}  Video scale: 0.1 to 1\par{}  Frames per second: 15 to 60\par{}  Sample rate: 8000, 11025, 22050, 44100, 48000\par{}  Video codec: MPEG-4, H-264, H-265, VP8\par{}  Audio codec: MPEG, Vorbis, Opus\par{}  Multiplexer: MPEG-TS, ASF/WMV, MKV, Ogg/Ogm, Webm \par{} Network interference: low, mid, high & 180 per protocol \\
\bottomrule
\end{tabularx}
\end{table*}
 
\begin{table*}[t]
\centering
\caption{Comprehensive summary of the tools, parameters, and methods used to generate malicious traffic.}\label{tab:mirai_params1}.\label{tab:attack_params}
\begin{tabularx}{\textwidth}{@{}lXX>{\hsize=7cm}X>{\hsize=1.75cm}X@{}}
\toprule
\textbf{Traffic Type} & \textbf{Traffic Subtype} & \textbf{Tool} & \textbf{Parameters Varied} & \textbf{Combinations} \\ 
\midrule
Bruteforce 
 & FTP & Patator, Filezilla & Network interference: low, high & 1 \\ 
 & \ac{DNS} (Fwd, Rev) & Patator, dnsmasq & & 2 \\ 
 & SSH & Patator &  & 1 \\ 
 & Telnet & Patator &  & 1 \\ \cmidrule{2-5}
 & HTTP fuzz & Patator, Apache, PhpMyAdmin &  Request method: GET, POST \par{} Network interference: low, high & 2  \\ 
\midrule
DoS 
 & MAC & macof & Network interference: low, high & 1 \\ \cmidrule{2-5}
 & HTTP & GoldenEye & Request method: GET, POST, Random\par{} Number of concurrent workers: 1, 20, 50\par{} Number of concurrent sockets: 100, 500, 1000 \par{} Network interference: low, high & 27 \\ \cmidrule{2-5}
 & \ac{ICMP} & Hping3 &   Payload size: 50, 500, 5000, 50000 bytes \par{}  Speed of pkt send: fast, faster, flood \par{} Network interference: low, high & 16 \\ \cmidrule{2-5}
 & UDP & Hping3 &   Payload size: 50, 500, 5000, 50000 bytes\par{}  Speed of pkt send: fast, faster, flood\par{}  Random source port\par{}  Bad UDP checksum (boolean) \par{} Network interference: low, high & 24 \\ \cmidrule{2-5}
 & TCP ACK \par{} TCP CWR \par{} TCP ECN \par{} TCP FIN \par{} TCP PSH \par{} TCP RST \par{} TCP SYN \par{} TCP URG & Hping3 & Speed of pkt send: fast, faster, flood\par{} Random source port\par{} Payload size: 50, 500, 5000, 50000 bytes\par{} Bad TCP checksum (boolean)\par{} TCP window size: default, 50, 1000\par{} Fake tcp data offset: 0, 5, 10 \par{} Network interference: low, high & 24 per attack\\ 
\midrule
Information Gathering 
 & Port Scan & Hping3 &  TCP Flags: ACK, FIN, PSH, RST, SYN, URG\par{}  Ports 1-65535 \par{} Network interference: low, high & 6 \\ \cmidrule{2-5}
 & TCP Port Scan & Nmap & Seven Scans: Connect, FIN, Maimon, NULL,  SYN/ACK, Window, Xmas \par{} Timing template: "Aggressive"\par{}  Bad checksum\par{}  Ports 1-65535\par{}  Send string as a payload\par{}  Payload size: 0, 50, 100, 5000 \par{} Network interference: low, high & 42 \\ 
\cmidrule{2-5}
 & UDP Port Scan & Nmap & Timing template: "Aggressive"\par{}  Bad checksum\par{}  Ports 1-65535\par{} Send string as a payload\par{}  Payload size: 0, 50, 100, 5000 \par{} Network interference: low, high & 6 \\ \cmidrule{2-5}
 & OS detection, version detection, script scanning, and traceroute & Nmap & Random MAC address \par{} Limit OS detection to only most likely matches\par{} Guess OS instead of relying on fingerprint matching \par{} Timing template: "Normal", "Aggressive"\par{} Bad checksum \par{} Set max-rate to 2 pps\par{}  Ports 1-65535\par{}  Send hexadecimal value as data payload\par{} Scan 100 most common ports\par{}  Payload size: 0, 50, 100, 5000\par{}  Enable fragmented IP packets \par{} Network interference: low, high & 6 \\ 
\cmidrule{2-5}
 & Ping Scan & Nmap & Seven Scans: ICMP echo, ICMP netmask request, ICMP timestamp request, SCTP INIT, TCP ACK, TCP SYN, UDP \par{} Timing template: "Normal", "Aggressive"\par{}  Bad checksum\par{}  Ports 1-65535\par{}  Send string as a payload\par{}  Payload size: 0, 50, 100, 5000 \par{} Network interference: low, high & 42 \\ 

\bottomrule
\end{tabularx}
\end{table*}
 

 
\subsection{Bidirectional Network Flow}\label{sec:bidirectional_flow}
A bidirectional network flow, or a session or conversation, exchanges network flows between two endpoints. Let $\mathcal{C}$ represent a bidirectional network flow composed of two individual network flows: $\mathcal{F}_1 = (p_1, p_2, \dots, p_m)$ and $\mathcal{F}_2 = (q_1, q_2, \dots, q_n)$. A session $\mathcal{C}$ is defined as a tuple $\mathcal{C} = (\mathcal{F}_1, \mathcal{F}_2)$, satisfying the following conditions: for every packet $p_i$ in $\mathcal{F}_1$ and every packet $q_j$ in $\mathcal{F}_2$, we have $p_i = (s_{\text{ip}}, s_{\text{prt}}, d_{\text{ip}}, d_{\text{prt}}, \tau)$ and $q_j = (d_{\text{ip}}, d_{\text{prt}}, s_{\text{ip}}, s_{\text{prt}}, \tau)$. This ensures that the network flows within the session are bidirectional, where one flow contains packets moving from the source to the destination, and the other flow contains packets moving from the destination back to the source.
 
\subsection{Intrusion Detection} 
Based on the definitions provided for unidirectional, bidirectional network flows, and subflows, we present the task of Intrusion Detection. This task involves classifying a network flow $\mathcal{F}$, whether unidirectional, bidirectional, or a subflow, into one of two categories: benign or malicious. For this purpose, we formally introduce the Intrusion Detection function $D: \mathcal{F} \rightarrow \{0, 1\}$ that assigns binary labels to network flows. In this mapping, an output of 0 corresponds to a benign flow, while an output of 1 signifies a malicious flow. Note that, in certain instances, the labels assigned may vary such that -1 represents malicious flows, and 1 represents benign flows. Function $D$ is learned from a training dataset of network flows $\mathcal{D} = \{(\mathcal{F}_1, y_1), (\mathcal{F}_2, y_2), \ldots, (\mathcal{F}_n, y_n)\}$, where each $\mathcal{F}_i$ is a network flow, and $y_i \in \{0, 1\}$ is the associated ground truth label. The Intrusion Detection problem can also be extended to address multi-class problems. In such scenarios, the function $D$ identifies whether a flow is benign or malicious and discerns the specific type or subtype of the traffic. Consequently, function $D$ is defined as $D: \mathcal{F} \rightarrow \{0, 1, 2, \ldots, k\}$. In this mapping, the output $k$ represents $k$ distinct classes of network traffic types or subtypes. Like the binary case, function $D$ is learned from a training dataset of network flows where each flow is linked to a label indicating its traffic type or subtype. Whether a binary or multi-class case, the optimal Intrusion Detection function accurately classifies unseen network flows, thereby contributing to identifying and mitigating potential network threats.
 
\section{Network Traffic Generation Details}\label{sec:appendix_diversification}
This section delves into the finer intricacies of our traffic generation procedures, detailing the specifications for each traffic type and the parameters that underwent variation during the generation process. Tables~\ref{tab:ben_params}, \ref{tab:attack_params}, and \ref{tab:mirai_params} together provide an extensive breakdown of the elements, including traffic types, subtypes, tools, varied parameters, and an estimated number of combinations. The ``Combinations" column indicates the count of traffic variations within a specific traffic subtype. This count is approximated based on the number of traffic permutations generated using the parameters varied unique to that subtype. For the Mirai botnet attack is represented in Table~\ref{tab:mirai_params}, with the ``Tool" column omitted since all derived attacks are associated with the Mirai botnet. We outline the parameters manipulated for each traffic subtype and their respective values and subsequently enumerate the varied parameters in the traffic generation process. 

In terms of benign traffic, as presented in Section~\ref{sec:ben_traffic}, we generated audio, background, text, and video traffic. For video traffic, we manipulated eleven parameters such as network interference levels (low, mid, high), video resolutions ranging from 240 to 1080 pixels, various audio and video bitrates, video scaling factors, frame rates, sample rates, an array of video codecs (e.g., MPEG-4, H-264), audio codecs (e.g., MPEG, Vorbis), and multiplexer types (e.g., MPEG-TS, MKV), presented in Table~\ref{tab:ben_params}. A compatibility-maintaining mapping was designed to synchronize video, audio, and multiplexer types interactions. During the data capture phase, we performed numerous rounds of traffic generation, with a Python script employed to facilitate \ac{VLC} video streaming. Specific considerations were also given to factors like audio and text traffic message length, and a deliberate 5\% client disconnection rate was introduced. The intricate manipulation of these parameters across various benign traffic subtypes was designed to capture real-world benign network traffic complexities.

To provide comprehensive coverage concerning traffic diversity, we conducted an intensive examination of various parameters within malicious traffic, attempting to vary the parameters extensively to achieve maximum coverage. Using dedicated tools listed in Table~\ref{tab:attack_params}, we systematically manipulated different types of attacks, such as Bruteforce and \ac{DoS}, carefully evaluating and altering the parameters specific to each attack. For all malicious traffic capture, apart from botnet traffic, we varied the network interference to capture data in low- and high-interference environments, expanded upon in Section~\ref{sec:testbed}. In \ac{DoS} attacks, we explored MAC, HTTP, ICMP, TCP, and UDP subtypes whilst altering attack-related parameters~\cite{alkasassbeh2016DDoS}. The packet size and speed of transmission, critical characteristics of DoS attacks, were also purposefully adjusted, incorporating transmission modes from Hping3, ``fast", ``faster", and ``flood" with 10 pps, 100pps, and over 1000pps respectively, and payload sizes ranging from 50 to 50,000 bytes. Furthermore, our approach included employing Nmap for conducting comprehensive scans, including OS detection, version detection, script scanning, and traceroute. The exploration encompassed a variety of configurations, enhancing the assessment of victim systems~\ref{tab:attack_params}. Some attacks, such as the DoS MAC flood, had limited variability due to the tools' constraints.

Finally, Table~\ref{tab:mirai_params} provides nuances of parameters within the Mirai Botnet attack context. We executed eight distinct attack vectors, each with unique optional manipulation parameters. A script was devised for Mirai DDoS attacks incorporating the varied parameters. Different subtypes of attacks, such as DDoS ACK and DDoS SYN, entailed specific manipulations like payload size, type of service, and random source/destination ports, culminating in multiple variations. This systematic diversification across each attack subtype contributes to our dataset's comprehensive and intricate representation of malicious network activities.
 
\begin{table*}[t]
\centering
\caption{Comprehensive summary of the tools, parameters, and methods used to generate Mirai Botnet traffic type.}\label{tab:mirai_params}
\begin{tabularx}{\textwidth}{@{}l>{\hsize=3.5cm}XX>{\hsize=1.75cm}X@{}}
\toprule
\textbf{Traffic Type} & \textbf{Traffic Subtype}& \textbf{Parameters Varied} & \textbf{Combinations} \\
\midrule
Mirai Botnet 
 & Scanning and Bruteforce & - & 1 \\ 
\cmidrule{2-4}
 & DDoS ACK &   Payload size: 50, 500, 1000 bytes\par{}  Type of service: none, 1\par{}  Random source and destination ports & 3 \\ 
\cmidrule{2-4}
 & DDoS SYN &   Flag: SYN, SYN URG, SYN PSH, SYN RST, SYN FIN, SYN ACK\par{}  Type of service: none, 1\par{}  Random source and destination ports & 3 \\ 
\cmidrule{2-4}
 & DDoS DNS &   Random source port & 1 \\ 
\cmidrule{2-4}
 & DDoS GREETH &   Payload size: 50, 500, 1000\par{}  Type of service: 6, 10, 70, 200\par{}  GCIP flag (boolean) & 3 \\ 
\cmidrule{2-4}
 & DDoS GREIP &   Payload size: 50, 500, 1000\par{}  Type of service: none, 1\par{}  Random source and destination ports\par{}  GCIP flag (boolean) & 4 \\ 
\cmidrule{2-4}
 & DDoS HTTP &  Request method: GET, POST\par{}  Number of connections: 50, 200, 800 & 2 \\ 
\cmidrule{2-4}
 & DDoS UDP &  Payload size: 50, 500, 1000\par{}  Type of service: none, 1\par{}  Random source and destination ports & 3 \\ 
\cmidrule{2-4}
 & DDoS UDP Plain &  Payload size: 50, 500, 1000\par{}  Random destination ports & 2 \\
\bottomrule
\end{tabularx}
\end{table*} 
 
\begin{table}[t]
    \centering
    \caption{XGBoost precision, recall, and F1 score for benign and malicious traffic.}
    \begin{tabularx}{\columnwidth}{lCCC} 
        \toprule
        \textbf{Category} & \textbf{Precision} & \textbf{Recall} & \textbf{F1 Score} \\
        \midrule
        Benign & 99.59 & 95.67 & 97.59 \\
        Malicious & 100.00 &  100.00 & 100.00 \\    
        \bottomrule
    \end{tabularx} 
    \label{tab:exp_sup_binary_detailed}
\end{table}
 
\begin{table}[t]
    \centering
    \caption{XGBoost precision, recall, and F1 score for each traffic \textit{type}.}
    \begin{tabularx}{\columnwidth}{l l CCC} 
        \toprule
            {} & \textbf{Traffic Type }& \textbf{Precision}  &       \textbf{Recall}  &   \textbf{F1 Score} \\
        \midrule
            \multirow{4}{*}{\rotatebox[origin=c]{90}{Benign}} & Audio &  97.30 &  100 &   98.63 \\
            {} & Background & 100 &   83.33 &   90.91 \\
            {} & Text &  97.22 &   87.50 &   92.11 \\
            {} & Video & 100 &   95.35 &   97.62 \\
        \midrule
            \multirow{4}{*}{\rotatebox[origin=c]{90}{Malicious}} & Bruteforce &  99.87 &   99.58 &   99.72 \\
            {} & DoS &  99.99 &  100 &   99.99 \\
            {} & Information Gathering & 100 &  100 &  100 \\
            {} & Mirai &  99.75 &   99.33 &   99.54 \\       
        \bottomrule
    \end{tabularx} 
    \label{tab:exp_sup_multiclass_type_detailed}
\end{table}  

\begin{table}[t]
    \centering
    \caption{Extra Trees precision, recall, and F1 score for each traffic \textit{subtype}.}
    \begin{tabularx}{\columnwidth}{l l CCC} 
        \toprule
        {} & \textbf{Traffic Subtype }& \textbf{Precision}  &       \textbf{Recall}  &   \textbf{F1 Score} \\
        \midrule
        \multirow{6}{*}{\rotatebox[origin=c]{90}{Benign}} & Audio & 94.74 &  100.00 &     97.30 \\
        {} & Background              &     100.00 &   83.33 &     90.91 \\
        {} & Text                    &      94.87 &   92.50 &     93.67 \\
        {} & Video HTTP              &      94.44 &   93.15 &     93.79 \\
        {} & Video RTP               &     100.00 &   97.14 &     98.55 \\
        {} & Video UDP               &      96.67 &  100.00 &     98.31 \\
        \midrule
        \multirow{26}{*}{\rotatebox[origin=c]{90}{Malicious}} & Bruteforce \ac{DNS}          &     100.00 &  100.00 &    100.00 \\ 
        {} & Bruteforce FTP          &     100.00 &   99.57 &     99.78 \\
        {} & Bruteforce HTTP         &     100.00 &   99.21 &     99.60 \\
        {} & Bruteforce SSH          &      99.37 &   99.75 &     99.55 \\
        {} & Bruteforce Telnet       &      98.02 &   96.51 &     97.26 \\
        {} & DoS ACK                 &      99.41 &   99.44 &     99.43 \\
        {} & DoS CWR                 &     100.00 &  100.00 &    100.00 \\
        {} & DoS ECN                 &     100.00 &  100.00 &    100.00 \\
        {} & DoS FIN                 &      99.49 &   99.47 &     99.48 \\
        {} & DoS HTTP                &      99.14 &   99.52 &     99.33 \\
        {} & DoS \ac{ICMP}                &     100.00 &  100.00 &    100.00 \\
        {} & DoS MAC                 &     100.00 &  100.00 &    100.00 \\
        {} & DoS PSH                 &      99.45 &   99.36 &     99.40 \\
        {} & DoS RST                 &      99.62 &   99.67 &     99.64 \\
        {} & DoS SYN                 &      99.99 &   99.98 &     99.99 \\
        {} & DoS UDP                 &      99.99 &  100.00 &    100.00 \\
        {} & DoS URG                 &     100.00 &  100.00 &    100.00 \\
        {} & Information Gathering   &     100.00 &   99.99 &    100.00 \\
        {} & Mirai DDoS ACK           &      99.87 &   99.33 &     99.60 \\
        {} & Mirai DDoS \ac{DNS}           &      99.99 &   99.98 &     99.99 \\
        {} & Mirai DDoS GREETH        &      44.44 &   50.00 &     47.06 \\
        {} & Mirai DDoS GREIP         &      27.27 &   30.00 &     28.57 \\
        {} & Mirai DDoS HTTP          &      95.95 &   93.61 &     94.77 \\
        {} & Mirai DDoS SYN           &      99.46 &   99.75 &     99.61 \\
        {} & Mirai DDoS UDP           &      58.33 &   50.00 &     53.85 \\
        {} & Mirai Scan and Bruteforce   &      97.96 &   98.21 &     98.09 \\  
        \bottomrule
    \end{tabularx} 
    \label{tab:exp_sup_multiclass_subtype_detailed}
\end{table}

\end{document}